\documentclass[twocolumn]{aastex631}

\begin{document}

\title{A Possible Non-Transiting Companion in the TOI-2109 System Supported by Transit Timing Variations}

\author[0009-0003-9593-7460]{Kaviya Parthasarathy}
\affiliation{Institute of Astronomy, National Tsing Hua University, Hsinchu 30013, Taiwan}
\email{kaviyasarathy1998@gmail.com}

\author[0009-0000-0807-2170]{Prangsutip Cherdwongsung}
\affiliation{Institute of Astronomy, National Tsing Hua University, Hsinchu 30013, Taiwan}
\email{prangsutip@gmail.com}

\author[0000-0001-7359-3300]{Ing-Guey Jiang}
\affiliation{Institute of Astronomy, National Tsing Hua University, Hsinchu 30013, Taiwan}
\affiliation{Department of Physics, National Tsing Hua University, Hsinchu 30013, Taiwan}
\email{jiang@phys.nthu.edu.tw}

\author[0000-0001-8677-0521]{Li-Chin Yeh}
\affiliation{Institute of Computational and Modeling Science, National Tsing Hua University, Hsinchu 30013, Taiwan}
\email{lichinyeh@mx.nthu.edu.tw}

\author{Hsin-Min Liu}
\affiliation{Institute of Astronomy, National Tsing Hua University, Hsinchu 30013, Taiwan}

\author[0009-0003-8446-4557]{Shraddha Biswas}
\affiliation{Institute of Astronomy, National Tsing Hua University, Hsinchu 30013, Taiwan}

\author[0000-0002-8988-8434]{D. Bisht}
\affiliation{Indian Centre For Space Physics, 
466, Barakhola, Singabari road, Netai Nagar, Kolkata, West Bengal, 700099, India}

\author[0000-0001-7234-7167]{Napaporn A-thano}
\affiliation{National Astronomical Research Institute of Thailand, 260 Moo 4, Donkaew, Mae Rim, Chiang Mai, 50180, Thailand}

\author[0000-0003-3251-3583]{Supachai Awiphan}
\affiliation{National Astronomical Research Institute of Thailand, 260 Moo 4, Donkaew, Mae Rim, Chiang Mai, 50180, Thailand}

\author{Vineet Kumar Mannaday}
\affiliation{
Department of Physics, Govt. Niranjan Kesharwani College, Kota, Bilaspur (C.G.)-495113, India}

\author{Parijat Thakur}
\affiliation{Department of Pure \& Applied Physics, Guru Ghasidas Vishwavidyalaya (A Central University), Bilaspur (C.G.) - 495 009, India}

\author{Evgeny Griv}
\affiliation{Department of Physics, Ben-Gurion University, Beer-Sheva 84105, Israel} 

\author{David Mkrtichian}
\affiliation{National Astronomical Research Institute of Thailand, 260 Moo 4, Donkaew, Mae Rim, Chiang Mai, 50180, Thailand}

%karen.collins@cfa.harvard.edu  (LCO scheduling, data reduction, time contribution)
\author[0000-0001-6588-9574]{Karen A.\ Collins}
\affiliation{Center for Astrophysics \textbar \ Harvard \& Smithsonian, 60 Garden Street, Cambridge, MA 02138, USA}
%emma.nabbie@cfa.harvard.edu
\author[0000-0003-0571-2245]{Emma Nabbie}
\affiliation{Center for Astrophysics \textbar \ Harvard \& Smithsonian, 60 Garden Street, Cambridge, MA 02138, USA}

%rpschwarz@comcast.net   (LCO data reduction)
\author[0000-0001-8227-1020]{Richard P. Schwarz}
\affiliation{Center for Astrophysics \textbar \ Harvard \& Smithsonian, 60 Garden Street, Cambridge, MA 02138, USA}

%thestockdalefamily@bigpond.com (LCO data reduction)
\author[0000-0003-2163-1437]{Chris Stockdale}
\affiliation{Hazelwood Observatory, Australia}

%khalid.barkaoui@uliege.be  (LCO data reduction)
\author[0000-0003-1464-9276]{Khalid Barkaoui}
\affiliation{Instituto de Astrof\'isica de Canarias (IAC), Calle V\'ia L\'actea s/n, 38200, La Laguna, Tenerife, Spain}
\affiliation{Astrobiology Research Unit, Universit\'e de Li\`ege, 19C All\'ee du 6 Ao\^ut, 4000 Li\`ege, Belgium}
\affiliation{Department of Earth, Atmospheric and Planetary Science, Massachusetts Institute of Technology, 77 Massachusetts Avenue, Cambridge, MA 02139, USA}

%narita@g.ecc.u-tokyo.ac.jp  (MuSCAT3/4 science team)
\author[0000-0001-8511-2981]{Norio Narita}
\affiliation{Komaba Institute for Science, The University of Tokyo, 3-8-1 Komaba, Meguro, Tokyo 153-8902, Japan}
\affiliation{Astrobiology Center, 2-21-1 Osawa, Mitaka, Tokyo 181-8588, Japan}
\affiliation{Instituto de Astrofisica de Canarias (IAC), 38205 La Laguna, Tenerife, Spain}

%afukui@g.ecc.u-tokyo.ac.jp   (MuSCAT3/4 science team)
\author[0000-0002-4909-5763]{Akihiko Fukui}
\affiliation{Komaba Institute for Science, The University of Tokyo, 3-8-1 Komaba, Meguro, Tokyo 153-8902, Japan}
\affiliation{Instituto de Astrofisica de Canarias (IAC), 38205 La Laguna, Tenerife, Spain}

%apopowicz@polsl.pl (SUTO observations and data reduction)
\author{Adam Popowicz}
\affiliation{Silesian University of Technology Observatory, Gliwice, Poland}

%nikola.antonov@iaps.institute (SW PDS250 observations and data reduction)
\author{Nikola Antonov}
\affiliation{Institute for Advanced Physical Studies, 111 Tsarigradsko Shose Blvd., Sofia, Bulgaria}

\begin{abstract}
Transit timing variations are a powerful method for detecting non-transiting planets, particularly in systems near mean-motion resonances. We investigate the ultra-short-period hot Jupiter TOI-2109 b using 87 transit epochs spanning two decades (2006-2025), combining observations from \textit{TESS}, \textit{CHEOPS}, \textit{HST}, and ground-based photometry.
We compare five timing models representing different physical scenarios: a linear ephemeris, orbital decay, apsidal precession, a sinusoidal model, and a two-planet $N-$ body dynamical model. The orbital-decay fit yields  $\dot{P} = -3.96 \pm 2.05$ ms yr$^{-1}$, providing no significant evidence for ongoing orbital decay. 
The sinusoidal model reveals a periodic 
transit-timing-variation signal with an amplitude of $\approx 44\,s$. Dynamical modelling favours an additional non-transiting companion near the 7:3 
mean-motion resonance with an orbital period of $1.57813\pm 8.9\times10^{-5}$ days and a mass of $0.70 \pm 0.02\,M_{\rm Jup}$. The two-planet model yields $\Delta \mathrm{BIC}=4.95$ relative to the sinusoidal model, indicating positive evidence that the observed transit timing variations are best explained by gravitational perturbations from an additional planet.
\end{abstract}


\section{Introduction} \label{sec:intro}
The discovery of thousands of exoplanets since the first detected exoplanet, 51 Pegasi b \citep{1995Natur.378..355M}, has transformed our understanding of planetary systems. These discoveries reveal a diversity of worlds and orbital configurations far beyond what is observed in the Solar System \citep{2015ARA&A..53..409W}. Studying these systems provides crucial insights into the physical and dynamical processes that shape planetary evolution \citep{2012A&A...547A.111M}.

Among the known population of close-in giant planets, ultra-hot Jupiters (UHJs), defined as planets with orbital periods shorter than one day ($P<1$ day), are among the most challenging systems to explain with current planet formation theories \citep{2018ARA&A..56..175D}. Only a handful of such systems are currently known, and their rarity, occurring around fewer than about 0.5\% of Sun-like stars, makes each one especially valuable for testing models of planetary evolution \citep{2006AcA....56....1G,2012ApJS..201...15H,2013ApJ...766...81F,2013MNRAS.433..889K,2018AJ....155...89P}. The extreme environments in which these planets reside make it unlikely that they formed in situ. As a result, UHJs provide important opportunities to investigate the mechanisms responsible for planetary migration and orbital evolution \citep{2019AJ....157..180P}.

TOI-2109 b stands as the most extreme known representative of this rare population, discovered by \citet{2021AJ....162..256W} using data from NASA's Transiting Exoplanet Survey Satellite (TESS). 
TOI-2109 b is an ultra-hot gas giant with a mass of approximately $5.02\,M_{\rm Jup}$  and a radius of approximately $1.35\,R_{\rm Jup}$, orbiting its host, a rapidly rotating F-type star ($v \sin i_\star = 81.9 \pm 1.7~{\rm km~s^{-1}}$), with a planetary orbital period of only $P \approx 0.6724$ days, or about 16 hours. The equilibrium temperature of TOI-2109 b exceeds $3500$ K, making it the second-hottest exoplanet known and placing it firmly in the class of UHJs, whose dayside atmospheres are thought to undergo thermal dissociation of molecular hydrogen \citep{2018ApJ...857L..20B}.

Since the orbital period of TOI-2109 b is shorter than the stellar rotation period, $P_{\rm rot}/\sin i_\star \approx 1.05$ day, tidal theory predicts that the planet transfers angular momentum to the host star and therefore gradually spirals inward through orbital decay \citep{1973ApJ...180..307C,1996Sci...274..954R,2020ApJ...888L...5Y,2025ApJ...988...66A}. 

The predicted orbital period decay rate, $dP/dt$ , for TOI-2109 b is among the largest known for hot Jupiter systems \citep{2021AJ....162..256W}, with estimated values ranging from about 10 to 740 ms\,yr$^{-1}$ depending on the assumed stellar tidal quality factor, $Q'_{\star}$, and the age of the host star. 
This makes TOI-2109 b an important system for studying the efficiency of stellar tidal dissipation mechanisms
including inertial wave excitation in convective envelopes and internal gravity wave propagation in radiative regions, both of which remain poorly constrained in exoplanet systems \citep{2025ApJ...988...66A,2007ApJ...661.1180O}.

Adding further intrigue to the system, \citet{Harre2024AA} identified low-amplitude transit-timing variations (TTVs) in TOI-2109~b. Although no additional transiting body was directly detected, the TTV signal is consistent with the gravitational influence of a close outer companion on a circular orbit with period $P_c \approx 1.57$ days. If confirmed, TOI-2109~b would join the rare class of hot Jupiter systems hosting close-in planetary companions.

The possible presence of a nearby companion to TOI-2109 b raises important questions about the dynamical structure of the system. In multi-planet systems, mean-motion resonances (MMRs), in which the orbital periods of two planets follow a simple integer ratio, play a key role in determining long-term orbital stability and driving TTVs. Deviations from exact resonance may constrain the strength of tidal dissipation over the system’s evolution \citep{2014AJ....147...32G}, and the resonant configuration itself can preserve information about the system’s formation and migration history \citep{1999ssd..book.....M,2012ApJ...761..122L,2015ApJ...802..116D}.

The origin of the observed TTVs in TOI-2109 b remains uncertain. Possible explanations include gravitational perturbations from an additional planet, apsidal precession of an eccentric orbit, or orbital decay driven by tidal dissipation, each of which can produce timing signals that are difficult to distinguish without sufficient temporal coverage and precision \citep{2016AJ....151...17J,2026AJ....171..321A}. A multi-model analysis that tests all of these scenarios is therefore necessary to disentangle their effects.

In this work, we apply five independent analytical and dynamical models: a linear ephemeris model, an orbital decay model, an apsidal precession model, a sinusoidal model, and a two-planet $N-$body model \citep{2024NewA..10602130Y,2025AJ....169..243J}. This combined approach allows us to consistently evaluate which physical process best explains the observed timing variations.

Previous studies of TOI-2109 b \citep{2021AJ....162..256W,Harre2024AA,2025ApJ...988...66A} were limited by the temporal baseline and the amount of available transit-timing data, leaving the origin of the TTVs and the presence of an additional companion unconfirmed. To improve the timing baseline and increase the number of available transit epochs, we combine ground-based observations with space-based photometric data from TESS, CHEOPS, and HST. Using this expanded dataset, we present updated transit mid-times and perform a comprehensive re-analysis of the TOI-2109 system.

This paper is organized as follows. Section \ref{sec:data} describes our observational dataset, data reduction procedures, and light curve fitting process. Section \ref{sec:ttvanalysis} presents the TTV analysis using the linear, orbital decay, apsidal precession, and sinusoidal models. Section \ref{sec:OSA} describes the orbital stability analysis. Section \ref{sec:DM} presents an identification of the MMR, differential evolution optimization, dynamical modeling of the TTVs, and model selection. Section \ref{sec:R&D} presents our conclusions.

\section{Observational data} \label{sec:data}
\subsection{Transiting Exoplanet Survey Satellite (TESS)}

TOI-2109 b was observed by TESS during Sectors 25, 52, and 79. Photometric data for all three sectors were obtained from the Science Processing Operations Center (SPOC) pipeline and downloaded using the \texttt{lightkurve} package \citep{2018ascl.soft12013L}. Sector 25 is available only at a 30-minute (1800\,s) cadence. Given that the transit duration of TOI-2109 b is approximately 1\,h 48\,min, the 30-minute cadence yields too few in-transit data points for precise transit modeling, and Sector 25 was therefore excluded from the analysis. For Sectors 52 and 79 we used the standard 2-minute (120\,s) cadence. 

\begin{figure*}[ht]
    \centering
    \includegraphics[width=0.8\linewidth, height=10cm]{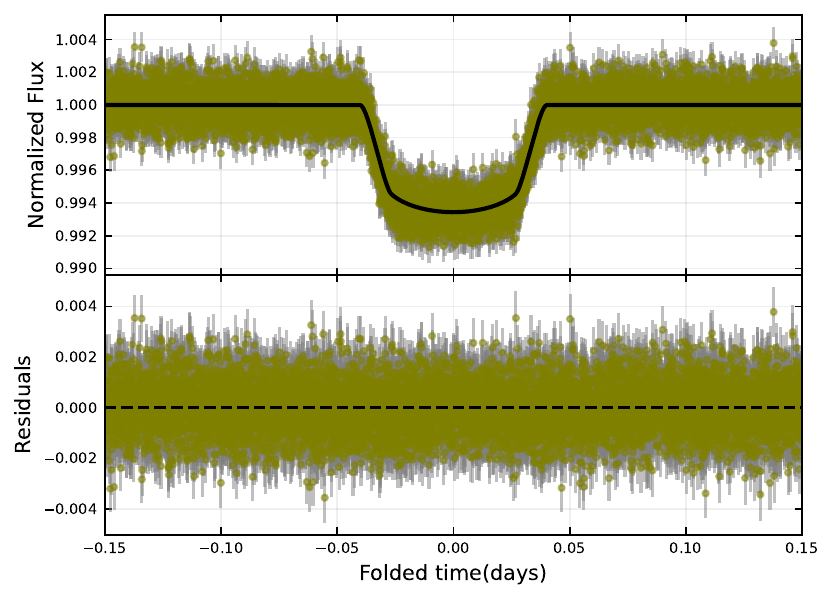}
    \caption{Phase-folded TESS transit light curve of TOI-2109 b, combining data from sectors 52 and 79}
    \label{fig:combined_TESS}
\end{figure*}

Since each of the remaining sectors contains multiple transit events, it was necessary to extract individual transits for the
TTV analysis. Using a reference mid-transit time ($T_0$) and orbital period ($P$) for TOI-2109b from \citet{2021AJ....162..256W}, the light curves from Sectors 52 and 79 were divided into independent transit events, yielding a total of 47 light curves used in this analysis. Sector 52 spans May 19 to June 12, 2022, while Sector 79 spans May 22 to June 18, 2024. The combined phase-folded light curves from these sectors are shown in Figure~\ref{fig:combined_TESS}.
%combined figure

\subsection{Characterising Exoplanet Satellite (CHEOPS)}
\begin{figure*}[ht]
    \centering
    \includegraphics[width=0.8\linewidth, height=10cm]{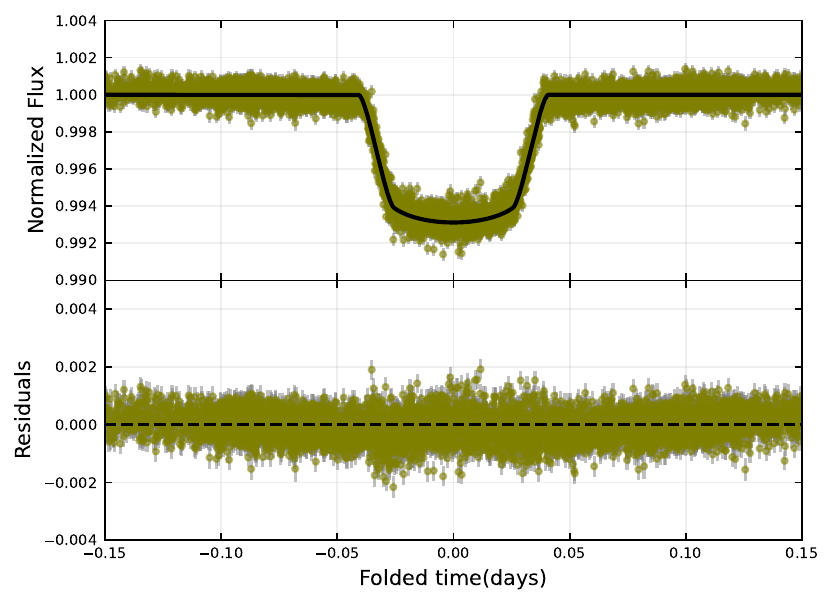}
    \caption{Phase-folded CHEOPS light curve of TOI-2109 b}
    \label{fig:combined_cheops}
\end{figure*}

CHEOPS photometric observations of TOI-2109 b were obtained from the CHEOPS archive\footnote{\url{https://cheops-archive.ssdc.asi.it/archive_browser/}}. These data are processed using the CHEOPS Data Reduction Pipeline (DRP), which provides aperture photometry across multiple aperture sizes in FITS format \citep{2020A&A...635A..24H}. We chose the one with the lowest RMS scatter, as a lower RMS indicates a more stable and less noisy light curve. Data points deviating by more than $3\sigma$ from the local baseline were identified and removed as outliers. Additionally, light curves with incomplete or insufficient in-transit coverage were discarded.
After applying these selection criteria, 25 CHEOPS light curves were retained for subsequent analysis. 
Before fitting the individual light curves, we corrected for instrumental variations associated with the spacecraft roll angle. To remove this roll-angle effect, we modeled the flux variations as a function of the roll angle $\phi$ using a three-harmonic Fourier series.
\begin{equation}
F_{\text{roll}}(\phi) = a_0 + \sum_{k=1}^{3} \left[ a_k \cos(k \phi) + b_k \sin(k \phi) \right]
\end{equation}
where $\phi$ is the roll angle, $a_k$ and $b_k$ are the harmonic coefficients. We fitted this three-harmonic model to each CHEOPS light curve using least squares. The observation log of the CHEOPS data is given in Table~\ref{tab:cheops_log}. The individual light curves are shown in Fig.~\ref{app:cheops_25}, whereas the combined phase-folded light curve is shown in Fig.~\ref{fig:combined_cheops}.

\subsection{Hubble Space Telescope (HST)}

Raw HST/WFC3 FITS files corresponding to a single transit observation of TOI-2109 b were retrieved from the 
Mikulski Archive for Space Telescopes (MAST) at the Space Telescope Science Institute\footnote{\url{https://exo.mast.stsci.edu/exomast_planet.html?planet=TOI2109b}}. The observation was carried out with the WFC3/IR instrument using the G141 grism, with an exposure time of 134\,s. The data were reduced using the \texttt{iraclis}\footnote{\url{https://github.com/ucl-exoplanets/Iraclis}} pipeline \citep{2016ApJ...832..202T}, which processed the raw exposure files and produced a white-light curve along with 25 spectrophotometric light curves in individual wavelength bins. The white-light transit data exhibited significantly larger timing uncertainties compared to the individual spectroscopic bins. We therefore used the 25 spectrophotometric bins to determine the mid-transit time for this epoch. To ensure a consistent fitting approach across all datasets, the time in BJD, flux, and associated uncertainties were extracted from the \texttt{iraclis} outputs and subsequently modeled using \texttt{Pylightcurve} (see Section~\ref{sec:fitting}). The mid-transit times derived from each of the 25 spectral bins were combined using a weighted average, resulting in a single HST mid-transit time adopted for the TTV analysis.
The light curves from the 25 spectrophotometric bins are shown in Appendix~\ref{app:25_hst}.

\subsection{TESS Follow-up Observing Program (TFOP)}

As part of TFOP, several ground-based observers contribute photometric follow-up observations of TESS Objects of Interest (TOIs). These light curves are made publicly available through the ExoFOP-TESS repository\footnote{\url{https://exofop.ipac.caltech.edu/tess/}}. 
Several of these observations were obtained with telescopes of the Las Cumbres 
Observatory Global Telescope (LCOGT) network \citep{Brown:2013} and the observation obtained on 2022 April 7 was carried 
out with the MuSCAT3 multi-band imager whose 
raw images are calibrated using the \texttt{BANZAI} pipeline \citep{McCully:2018}. Photometric extraction from these calibrated images, as well as from the other 
TFOP light curves, was performed by the individual observers using 
\texttt{AstroImageJ} \citep{2017AJ....153...77C}.

 Therefore, no additional data reduction was performed, and the pre-processed light curves were used directly in our analysis.

\begin{figure*}[ht]
    \centering
    \includegraphics[width=0.8\linewidth, height=12cm]{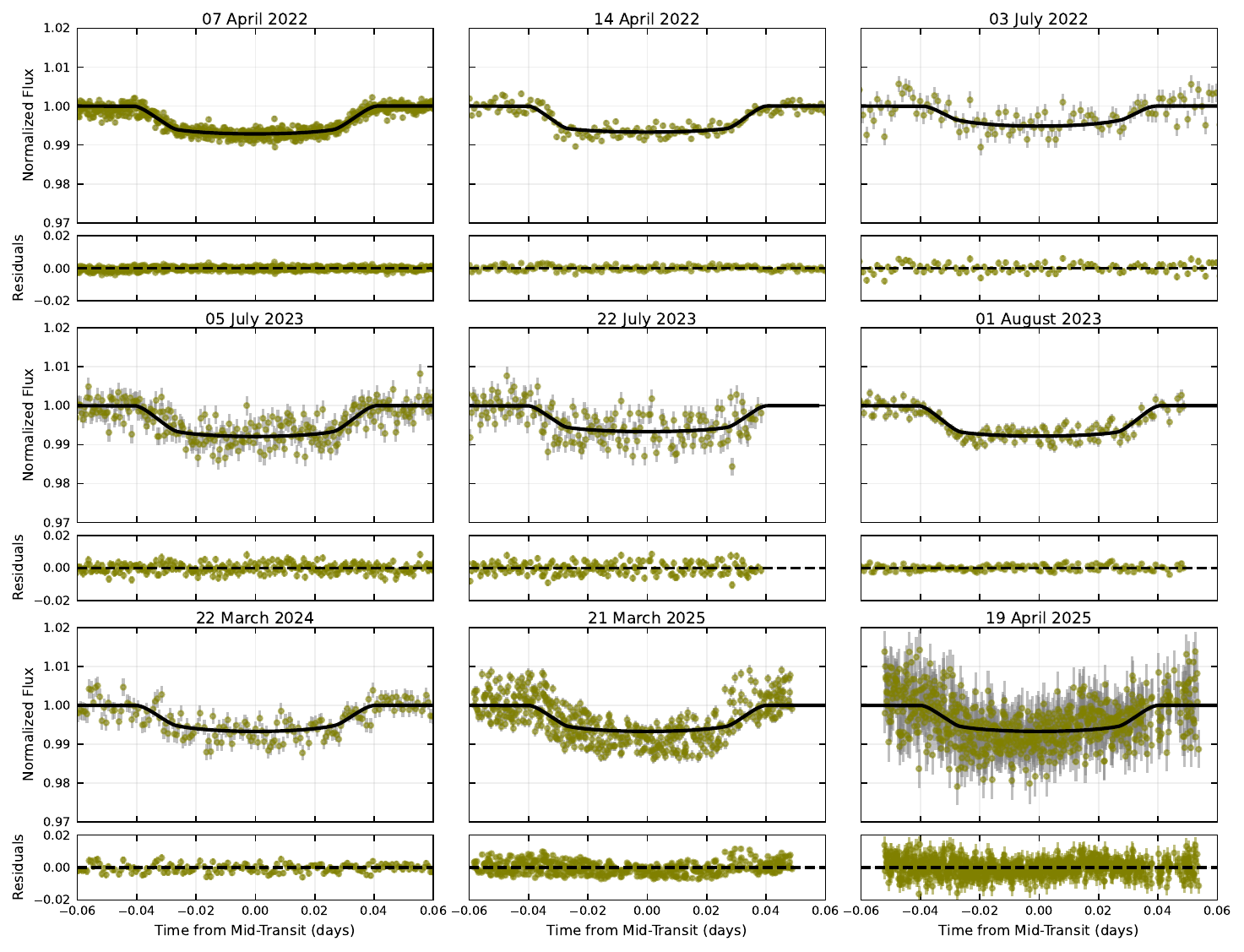}
    \caption{New photometric data used in our analysis. Green points show the observed data, and the black solid line shows the best-fit model obtained with \texttt{PyLightcurve}. The light curves from 21 March 2025 and 19 April 2025 were obtained by our ground-based observation, while the remaining light curves are from TFOP.}
    \label{fig:new_fitted_lightcurves}
\end{figure*}

A subset of the available TFOP light curves were previously used by \citet{2021AJ....162..256W}. 
We additionally include the remaining publicly available observations not incorporated in that study. When multiple observations existed for the same transit epoch (e.g., from different observers or filters), only the highest-quality light curve was retained, and the others were excluded to avoid duplicate timing measurements that could bias the transit timing analysis. In total, 15 TFOP light curves were used, of which 7 are new to this work, the corresponding new light curves are shown in Figure~\ref{fig:new_fitted_lightcurves}.

%new lightcurves

\subsection{Ground-based Observations} \label{sec:ground data}

We conducted two new photometric transit observations of TOI-2109 b using 1-m Thai National Observatory (TNO) telescope in Thailand, equipped with a 2K$\times$2K Andor iKon-L CCD camera. The complete log of the observations including HST,TFOP and two new ground-based  observations, is provided in Table~\ref{table:observation_log}. The raw images were calibrated using \texttt{AstroImageJ} \citep{2017AJ....153...77C}, which performed standard corrections for bias, dark current, and flat-field effects. Comparison stars of similar brightness to the target were carefully selected to perform differential aperture photometry. The resulting light curves consist of time in BJD, relative flux, and associated uncertainties, which were used for subsequent analysis, and their corresponding light curves are shown in Figure~\ref{fig:new_fitted_lightcurves}.
%The normalized fluxes and uncertainties are listed in Table~\ref{table:Norm.lightcurves} 
%table 1
%observation log
\begin{table*}
\centering
\caption{Observation log of the new TOI-2109b light curves used in this work.}
\label{table:observation_log}
\begin{tabular}{cccccc}
\hline\hline
Date & Epoch & Telescope & Filter &  Transit Coverage \\
\hline
07 April 2022 &444 & LCO-HAL & \textit{i'}&  Full\\
14 April 2022 &454 &LCO-CTIO & \textit{z'} & Full\\
03 July 2023 & 574&SUTO-Pyskowice & \textit{B}  &  Full\\
05 July 2023 & 1119 &LCO-HAL & \textit{}{z'} &  Full\\
22 July 2023 &1144 &LCO-McD& \textit{z'}& Full\\
1 August 2023 & 1160&LCO-CTIO
& \textit{z'}& Full\\
22 March 2024 &1508 &SW PDS250 & \textit{B} &Full\\
28 March 2024 & 1516& HST & \textit{G141 grism}& Full\\ 
21 March 2025  &2049 & TNO & \textit{R} &  Full \\
19 April 2025  &2092 & TNO &\textit{R} &  Full \\
\hline
\end{tabular}
\end{table*}
\subsection{Early WASP data}

In addition to the space-based, and the ground-based transit photometry, we incorporate additional archival photometry from the SuperWASP-N survey (Wide Angle Search for Planets; \citealp{2006PASP..118.1407P}), which observed TOI-2109 between 2006 and 2011. This dataset, phase-folded 
on the known ephemeris, was processed and analysed by \citet{Harre2024AA}, who confirmed 
the transit signal in the folded WASP light curve. We adopt this mid-transit time directly for our analysis rather than re-reducing the raw WASP photometry ourselves. Even TESS, with its multi-sector coverage since 2020, cannot match this early baseline, the WASP data extend our transit-timing coverage back by more than a decade beyond the earliest space-based observations, making it by far the longest-baseline dataset available for this system.

\subsection{Light Curve Fitting} \label{sec:fitting}

We corrected for stellar variability using a first-order polynomial detrending, following the formalism of \citet{2021AJ....162..256W}:

\begin{equation}
F_{\rm poly}(t) = \sum_{k=0}^{N} c_k (t-t_0)^k,
\end{equation}

where $c_k$ is the detrending coefficient of the $k$th-order term, $N$ is the polynomial order ($N=1$ in this work), and $t_0$ is a reference time. This correction was applied consistently to all light curves used in this work.

We performed the light curve fitting with the \texttt{PyLightcurve} package\footnote{\url{https://github.com/ucl-exoplanets/pylightcurve}} \citep{2016ApJ...832..202T}. The Initial stellar parameter values were adopted from the default inputs provided by the Exoplanet Characterisation Catalogue \citep{2022ExA....53..547K}, developed as part of the ExoClock project, and accessed via the \texttt{pylightcurve.get\_planet()} function. The initial planetary parameter values and their corresponding priors are listed in Table~\ref{table:Initial_parameters_TOI2109b}.

For the TESS, CHEOPS, and our ground-based observations, we performed a global transit fit to derive the system parameters. For each instrument, we combined the light curves and fitted them with \texttt{PyLightcurve}. The limb-darkening effect was modeled using a quadratic law, with corresponding coefficients $(u_1, u_2)$.
The free parameters were the  orbital period $(P)$ mid-transit time $(T_0)$, inclination $(i)$, scaled semi-major axis $(a/R_*)$, planet-to-star radius ratio $(R_p/R_*)$, and the limb-darkening coefficients $(u_1, u_2)$. The eccentricity $(e)$, and argument of periastron $(\omega)$ were held fixed. Uniform priors were applied to all free parameters. 
The parameters derived from the combined fit were fixed as the global transit-shape parameters when fitting individual light curves to measure their mid-transit times.
For the HST and TFOP data, each light curve was fitted independently. We applied the same priors as in the combined fit, except that the orbital period was held fixed.
Best-fit values and associated uncertainties were obtained using the \texttt{emcee} sampler \citep{2013PASP..125..306F}, as implemented by default within \texttt{PyLightcurve}. The resulting fitting parameters are presented in Table~\ref{table:newdata_results}, and the mid-transit times from all the light curves are provided in Table~\ref{app:midtimes_all}. The fitted light-curve data are available in machine-readable form; see Table~\ref{tab:flux_data}.

%Initial parameters

\begin{table*}
    \centering
    \caption{Initial values and priors of the planetary parameters used in the \texttt{PyLightcurve} fitting.}
    \label{table:Initial_parameters_TOI2109b}
    
    \begin{tabular}{cccc}
        \hline
        \hline
        Parameter & Initial value & Prior range & Prior distribution \\
        \hline
        \hline
        $P$ (days) & 0.672474 & [0.672464, 0.672484] & Uniform \\
        $T_0 - 2459000$ (BJD$_{\rm TDB}$) & 486.727 & [486.527, 486.927] & Uniform \\
        $i$ (deg) & 70.74 & [68.74, 72.74] & Uniform \\
        $a/R_\star$ & 2.268 & [2.16, 2.36] & Uniform \\
        $R_{\mathrm{p}}/R_\star$ & 0.0815 & [0.0715, 0.0915] & Uniform \\
        $e$ & 0.0 & -- & Fixed \\
        $\omega$ (deg) & 0.0 & -- & Fixed \\
        $u_1^{\dagger}$ & -- & [0, 1] & Uniform \\
        $u_2^{\dagger}$ & -- & [0, 1] & Uniform \\
        \hline
    \end{tabular}

    \footnotesize{
    $^{\dagger}${By default, \texttt{PyLightcurve} computes the initial limb-darkening coefficients with \texttt{ExoTETHyS} \citep{2020AJ....159...75M},
    \\based on the stellar parameters of the host star.}}
\end{table*}

%0-C figure
\begin{figure*}[ht]
    \centering
    \includegraphics[width=0.9\linewidth, height=10cm]{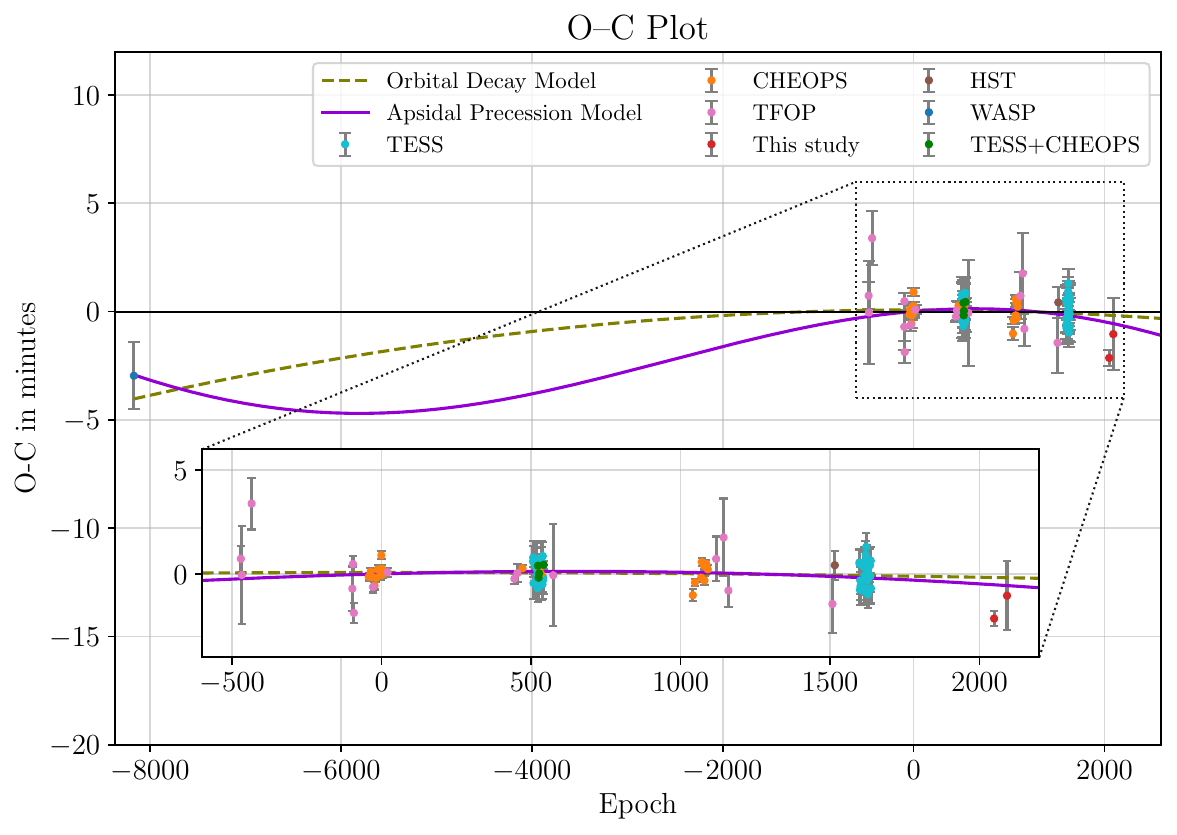}
    \caption{Transit Timing Variation (TTV) diagram for TOI-2109 b after subtracting the linear model. The plot compares the (O-C) values against linear, orbital decay, and apsidal precession models}.
     \label{fig:ocplot_TOI-2109b}
\end{figure*}

% fitting results
\begin{deluxetable*}{cccccccc}
\tabletypesize{\scriptsize}
\tablewidth{0pt}
\tablecaption{Fitting results for the new observations. \label{table:newdata_results}}
\tablehead{
\colhead{Date} & \colhead{P(days)} & \colhead{$T_{\mathrm{m}}$ or $T_0$} & \colhead{$a/R_\star$} & \colhead{$R_{\mathrm{p}}/R_\star$} & \colhead{$i$ ($^\circ$)} & \colhead{$u_1$} & \colhead{$u_2$}
}
\startdata
2022 April 07 & \nodata & $2459677.03745^{+0.00017}_{-0.00015}$ & $2.260^{+0.048}_{-0.012}$ & $0.0847^{+0.0011}_{-0.0009}$ & $70.72^{+0.83}_{-0.29}$ & $0.350^{+0.150}_{-0.190}$ & $0.157^{+0.383}_{-0.070}$ \\
2022 April 14 & \nodata & $2459683.76238^{+0.00036}_{-0.00024}$ & $2.267^{+0.077}_{-0.001}$ & $0.0815^{+0.0012}_{-0.0004}$ & $70.74^{+0.82}_{-0.22}$ & $0.291^{+0.150}_{-0.070}$ & $0.150^{+0.140}_{-0.080}$ \\
2022 July 03 & \nodata & $2459764.45920^{+0.00180}_{-0.00160}$ & $2.266^{+0.070}_{-0.032}$ & $0.0817^{+0.0040}_{-0.0040}$ & $70.72^{+0.63}_{-0.55}$ & $0.630^{+0.210}_{-0.320}$ & $0.156^{+0.398}_{-0.085}$ \\
2023 July 05 & \nodata & $2460130.95813^{+0.00090}_{-0.00060}$ & $2.268^{+0.051}_{-0.031}$ & $0.0887^{+0.0015}_{-0.0028}$ & $70.74^{+0.64}_{-0.56}$ & $0.360^{+0.200}_{-0.250}$ & $0.154^{+0.267}_{-0.082}$ \\
2023 July 22 & \nodata & $2460147.77070^{+0.00130}_{-0.00130}$ & $2.265^{+0.034}_{-0.067}$ & $0.0813^{+0.0053}_{-0.0041}$ & $70.76^{+0.76}_{-0.38}$ & $0.360^{+0.170}_{-0.220}$ & $0.151^{+0.213}_{-0.089}$ \\
2023 August 01 & \nodata & $2460158.52851^{+0.00058}_{-0.00051}$ & $2.268^{+0.024}_{-0.049}$ & $0.0881^{+0.0016}_{-0.0032}$ & $70.75^{+0.65}_{-0.49}$ & $0.290^{+0.090}_{-0.010}$ & $0.150^{+0.150}_{-0.100}$ \\
2024 March 22 & \nodata & $2460392.54905^{+0.00111}_{-0.00083}$ & $2.265^{+0.047}_{-0.060}$ & $0.0816^{+0.0041}_{-0.0041}$ & $70.74^{+0.72}_{-0.48}$ & $0.520^{+0.220}_{-0.280}$ & $0.161^{+0.464}_{-0.071}$ \\
\cutinhead{Combined fit}
TESS  & $0.672477^{+0.000067}_{-0.000033}$ & $2459378.45909^{+0.00009}_{-0.00007}$ & $2.268^{+0.086}_{-0.052}$ & $0.0808^{+0.0040}_{-0.0020}$ & $70.76^{+0.69}_{-1.04}$ & $0.400^{+0.120}_{-0.200}$ & $0.118^{+0.094}_{-0.056}$ \\
CHEOPS & $0.672475^{+0.000061}_{-0.000043}$ & $2459378.45916^{+0.00003}_{-0.00002}$ & $2.266^{+0.006}_{-0.053}$ & $0.0825^{+0.0015}_{-0.0037}$ & $70.79^{+0.19}_{-1.10}$ & $0.301^{+0.052}_{-0.075}$ & $0.108^{+0.034}_{-0.061}$ \\
TNO & $0.672474^{+0.000003}_{-0.000005}$ & $2459378.45662^{+0.00128}_{-0.00035}$ & $2.270^{+0.067}_{-0.003}$ & $0.0816^{+0.0041}_{-0.0036}$ & $70.73^{+0.28}_{-0.28}$ & $0.429^{+0.033}_{-0.212}$ & $0.154^{+0.101}_{-0.050}$ \\
\enddata
\tablecomments{$T_{\mathrm{m}}$ is the individual mid-transit time for each new observation; $T_0$ is the reference mid-transit time from the combined fits, for which the period was also allowed to vary.}
\end{deluxetable*}

\section{TTV Analysis} \label{sec:ttvanalysis}
To investigate potential TTVs in the TOI-2109 system, we tested three different models following the approach of \citet{2017AJ....154....4P,2025NewA..11902390P}. Each model represents a different physical mechanism that could cause deviations in the observed transit times from a strictly periodic signal.

\subsection{Linear Ephemeris Model}
The simplest model assumes a constant orbital period, where the mid-transit time varies linearly with epoch:

\begin{equation}
    T_m^c(E) = T_0 + EP,
\end{equation}
$ T_m^c(E)$ is the calculated mid-transit time at epoch (E), $T_0$ is the reference mid-transit time, and \textit{P} is the orbital period. This model serves as the baseline against which the other models are compared.
 
\subsection{Orbital Decay Model}
The second model accounts for orbital decay driven by tidal interactions between the star and the planet, where the orbital period decreases at a constant rate:

\begin{equation}
    T_d^c(E) = T_{d0} + E P_d + \frac{1}{2} \frac{dP_d}{dE}E^2 
\end{equation}

where $ T_d^c(E)$ is the calculated mid-transit time at epoch (E),$T_{d0}$ is the reference mid-transit time, $P_d$ is the orbital period, $ \frac{dP_d}{dE}$ is the period change per orbit.

\subsection{Apsidal Precession Model}
The third model considers the possibility that the planet follows a slightly eccentric orbit with a precessing argument of periastron \citep{2009IAUS..253..163W}.  Following \citet{1995Ap&SS.226...99G}, the mid-transit time is given by:
\begin{equation}
    T_{ap}^c(E) = T_{ap0} + P_sE- \frac{eP_scos(w_0 + E\frac{dw}{dE})}{\pi (1-\frac{\frac{dw}{dE}}{2\pi})},
\end{equation}
$  T_{ap0} $ is the reference mid-transit time, $ e $ is the eccentricity, $\mathbf{\omega_0}$ is the argument of periastron and $P_s $ is the sidereal period and $ \frac{dw}{dE}$ is the precession rate per orbit.

\subsection{Model Fitting}

Before fitting each of the models to the observed mid-transit times described above, we accounted for the uncertainties associated with our ground-based observations by adopting the likelihood formalism of \citet{2010arXiv1008.4686H}, we introduced an additional free parameter, $\ln f$. The resulting log-likelihood is given by,

\begin{equation}
\ln \mathcal{L} = -\frac{1}{2}\sum_i \left[\frac{(y_i-m_i)^2}{s_i^2}+\ln\left(2\pi s_i^2\right)\right],
\end{equation}

where $y_i$ and $m_i$ are the observed and predicted mid-transit times, 
respectively, and $s_i^2$ is the effective variance, given by

\begin{equation}
s_i^2 = \sigma_i^2 + f^2 .
\end{equation}

where $\sigma_i$ is the uncertainty on the observed mid-transit time, $f$ is an additional uncertainty (jitter), parameterized as $\ln f$ in the fit.

we used the \textit{emcee} Markov Chain Monte Carlo (MCMC) ensemble sampler \citep{2013PASP..125..306F}. 
For each model, 50 walkers were initialized and run for $2 \times 10^{4}$ steps, with the first $10^{3}$ steps discarded as burn-in, except for the apsidal model, which was run for $2 \times 10^{5}$ steps with a burn-in of $10^{4}$ steps. Convergence  was verified by ensuring that the chain length was at least 50 times the integrated autocorrelation time. Uniform priors were adopted for all free parameters. 

The best-fit values were taken as the maximum a posteriori (MAP) estimates. We adopted the MAP estimate rather than the posterior mean or median because it identifies the single most probable parameter value given the data, which is especially useful when the posterior distribution is asymmetric or non-Gaussian. The uncertainties were taken as the standard deviations of the posterior distributions. 
The priors and best-fitting values of the free parameters for the three models are listed in Table~\ref{table:3model_results}. The corresponding corner plots are shown in Appendix ~\ref{fig:Linearmodel_cornor}, \ref{fig:decaymodel_cornor}, and \ref{fig:Apsidalmodel_cornor}, while the observed minus calculated (O-C) diagrams are presented in Figure~\ref{fig:ocplot_TOI-2109b}.

The fitting results for the linear, orbital decay, and apsidal precession models yield 
$\chi_{\nu}^2$ values of 2.34, 2.25, and 2.18, and BIC values of 210.15, 205.12, and 
203.88, respectively.
The linear model, which shows the largest $\chi_{\nu}^2$ and 
BIC among the three, exhibits deviations that indicate the presence of TTVs in the system.

For the orbital decay model, we quantified the rate of orbital decay by estimating the 
modified stellar tidal quality factor, $Q_\star'$, which describes how efficiently the star dissipates tidal energy. Following \citet{2018AcA....68..371M}, $ Q_\star'$ can be expressed as:

\begin{equation}
Q_\star' = -\frac{27}{2}\pi\left(\frac{M_p}{M_\star}\right)\left(\frac{a}{R_\star}\right)^{-5}\left(\frac{dP_d}{dE}\right)^{-1}P,
\end{equation}

where $M_p$, $M_\star$, and $a/R_\star$ are adopted from \citet{2021AJ....162..256W}, and $dP_d/dE$ is the decay rate derived from our orbital-decay model fit.
%($\dot{P} = -2.08 \pm 1.39$ ms\,yr$^{-1}$). 

We obtain $Q_\star' \sim 1.85\times10^{7}$. This value lies toward the upper end of the range typically inferred for hot Jupiter host stars, $Q_\star' \sim 10^{5}$--$10^{7}$ \citep{2011A&A...529A..50L}, indicating comparatively weak tidal dissipation in the TOI-2109 system, falling below the
decay rates predicted by \citet{2021AJ....162..256W}, disfavoring rapid
tidal in-spiral as the current dominant process in this system.

The apsidal precession model yields the lowest $\chi_{\nu}^2$ and BIC, suggesting that the observed timing variations are likely influenced by gravitational interactions within the system.
%, with $\Delta\mathrm{BIC}_{\mathrm{apsidal-linear}} = 6.57$ and $\Delta\mathrm{BIC}_{\mathrm{apsidal-decay}} = 2.22$. 
However, these $\chi_{\nu}^2$ values remain close 
to one another, indicating that the current data do not strongly distinguish the models. Additionally, the O-C diagram exhibits a sinusoidal-like variation, motivating us also to test a sinusoidal model to explain the observed TTVs.

\begin{table*}
\centering
\small
\caption{The Best-fit Parameters for TOI-2109 b}
\label{table:3model_results}
\begin{tabular}{ccc}
\hline
\hline
Models & Uniform Priors & Best-fit Values \\
\hline
\hline
\textit{Linear Model} & & \\
$P$ [days] & [0.66, 0.68] & $0.67247409 \pm 6.5\times10^{-8}$ \\
$T_0$ [$\mathrm{BJD_{TDB}}-2450000$] & [9377, 9379] & $9378.45911 \pm 6.3\times10^{-5}$ \\
\hline
\textit{Orbital Decay Model} & & \\
$P$ [days] & [0.66, 0.68] & $0.67247407 \pm 6.5\times10^{-8}$ \\
$T_0$ [$\mathrm{BJD_{TDB}}-2450000$] & [9377, 9379] & $9378.459163 \pm 6.6\times10^{-5}$ \\
$dP/dE$ [days/orbit] & [-0.5, 0.5] & $(-8.5 \pm 4.4)\times10^{-11}$ \\
\hline
\textit{Apsidal Model} & & \\
$P$ [days] & [0.66, 0.68] & $0.6724736 \pm 2.15\times10^{-7}$ \\
$T_0$ [$\mathrm{BJD_{TDB}}-2450000$] & [9378.3, 9378.5] & $9378.45644 \pm 7.8\times10^{-4}$ \\
$e$ & [0.00055, 0.015] & $0.013 \pm 0.003$ \\
$\omega_0$ [rad] & [1.0, 5.0] & $2.58 \pm 0.66$ \\
$d\omega/dE$ [rad epoch$^{-1}$] & [0.00010, 0.0009] & $0.00039 \pm 0.00020$ \\
\hline
\end{tabular}
\end{table*}

\subsection{Sinusoidal Model\label{sinu_model}}
\begin{figure*}[ht]
    
    \includegraphics[width=0.9\textwidth, height=10cm]{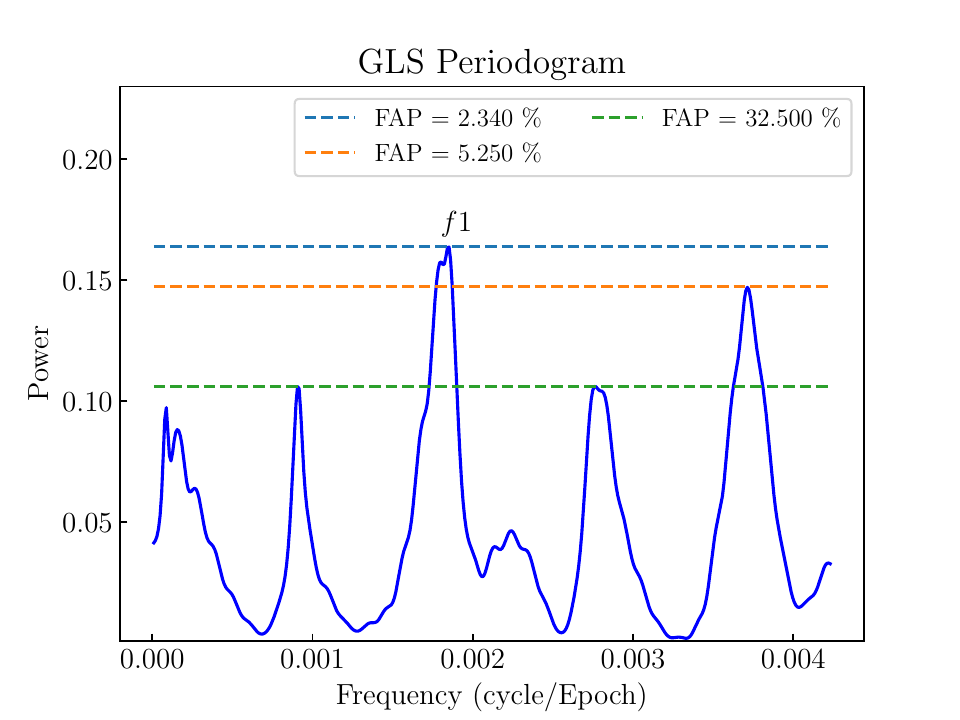}
    \caption{The blue solid lines indicate the periodic signals of the perturber, and the dashed lines indicate the corresponding false-alarm-probability (FAP) levels.}
     \label{fig:GLS}
\end{figure*}

\begin{figure*}
    \centering
    \includegraphics[width=0.9\textwidth, height=10cm]{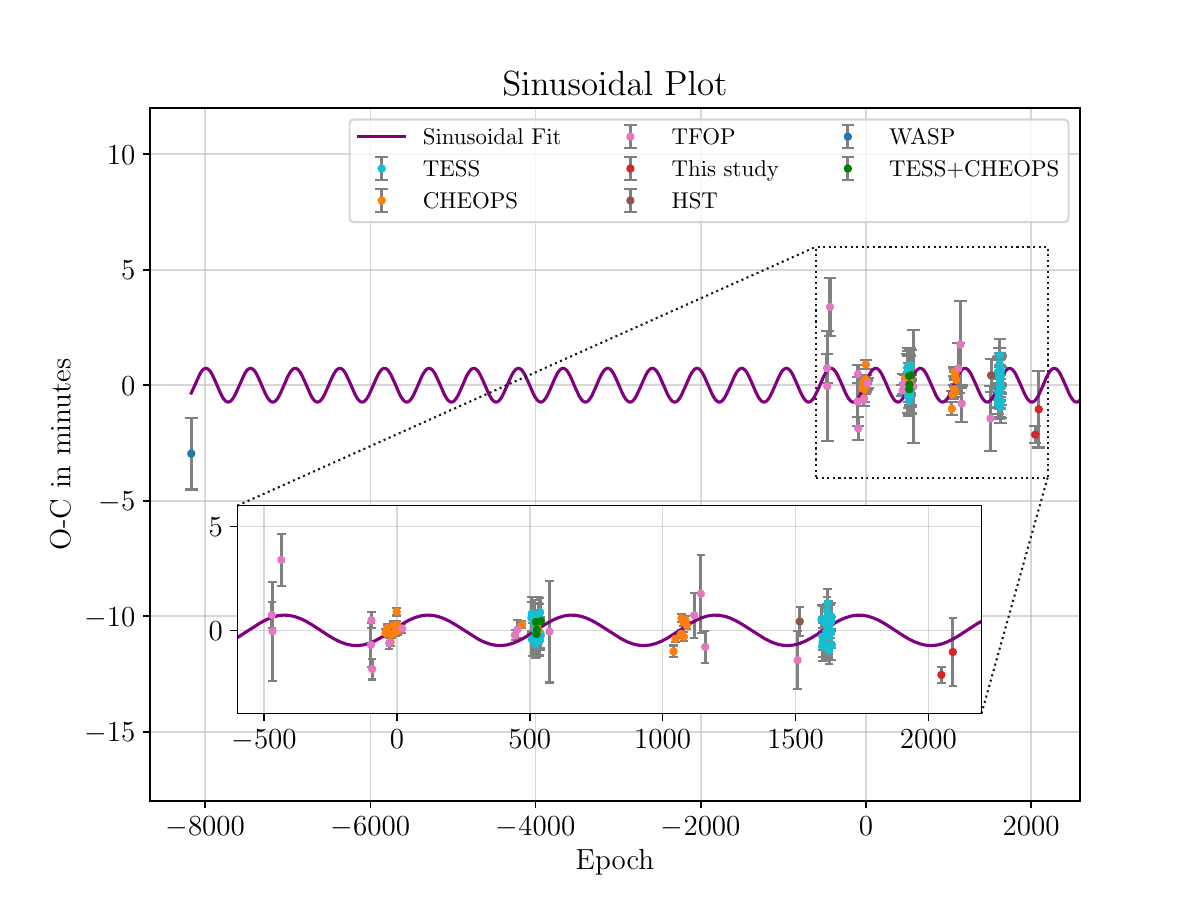}
    \caption{Sinusoidal model fitting for the most significant frequency $f1$ identified in the GLS periodogram analysis.
     \label{fig:sinusoidal_model}}
\end{figure*}

The fourth model fits the TTVs as a sinusoidal function of epoch, commonly used to search for periodic signals caused by gravitational perturbations from an unseen companion \citep{2019A&A...628A.116V}:

\begin{equation}
    \mathrm{TTV}(E) = A_{\mathrm{TTV}} \sin\!\left( 2\pi f E - \phi \right)
\end{equation}

where $A_{\mathrm{TTV}}$ is the amplitude of the TTV signal, $f$ is the frequency and $\phi$ is the phase. 
To identify the dominant frequency in the TTV signal, we applied the Generalized Lomb-Scargle (GLS) periodogram \citep{1976Ap&SS..39..447L,1982ApJ...263..835S} from PyAstronomy
\footnote{\url{https://pyastronomy.readthedocs.io/en/latest/pyTimingDoc/pyPeriodDoc/gls.html}} to the (O-C) residuals from the linear ephemeris fit, shown in Figure~\ref{fig:GLS}. The best-fit frequency was selected based on the lowest false alarm probability (FAP) of 2.34\%, which quantifies the probability that a detected peak arises from noise rather than a true periodic signal. The amplitude $A_{\mathrm{TTV}}$, frequency $f$, and phase $\phi$ were then fitted simultaneously as free parameters using the same MCMC approach described in Section~\ref{sec:ttvanalysis}. The dominant frequency was found to be f=0.00184 (cycle/Epoch), corresponding to a sinusoidal period of $\approx $ 540 days, with a TTV amplitude of $\approx$ 44\,s. The priors and best-fit values are listed in Table~\ref{table:frequency_2}. The best-fit sinusoidal model and the corresponding corner plot are shown in Figures~\ref{fig:sinusoidal_model} and~\ref{fig:sinusoidal_cornor}, respectively. The presence of a periodic signal motivates a search for an additional companion. We therefore next test dynamical stability.

%Gls

%sinusoidal model

%sinusoidal results
\begin{table*}
\centering
\caption{Best-fitting parameters of the sinusoidal model for the periodic signal ($f1$)}
\label{table:frequency_2}
\begin{tabular}{ccccc}
\hline
\hline
Sinisoidal Model& Uniform Priors & Best-Fit Value\\
\hline
\hline
$A_{\rm TTV}$ [min]&[0,2.0]&$ 0.73 \pm 0.20 $\\
f[cycle/Epoch]&[0.00155,0.00195]&$ 0.001849 \pm0.000032 $\\
$\phi$[rad]&[-0.78,-0.28]&$-0.20 \pm 0.17$\\
\hline
\end{tabular}
\end{table*}

\section{Orbital dynamics analysis} \label{sec:OSA}

To investigate the orbital dynamics of the TOI-2109 system and identify regions where an additional companion could remain in a regular mean motion resonance (MMR) with TOI-2109 b, we constructed a Mean Exponential Growth factor of Nearby Orbits (MEGNO) map using the N-body integration package \texttt{REBOUND} \citep{2012A&A...537A.128R}.
MEGNO is a chaos indicator; values converging to $ \langle Y \rangle \approx 2 $ indicate regular motion, while values significantly exceeding 2 indicate chaotic motion \citep{2000A&AS..147..205C}.
We employed the \texttt{WHFast} symplectic integrator \citep{2015MNRAS.452..376R}, which is well-suited for near-Keplerian systems due to its computational efficiency and long-term energy conservation properties.
 
A hypothetical perturbing body was introduced on a coplanar $(i_b=i_c)$, circular orbit $(e_b=e_c=0)$. Its orbital period was varied along the x-axis from 0.8 to 5 days, while its mass was varied along the y-axis from sub-Earth to super-Jovian values on a logarithmic scale, allowing us to probe a wide dynamic range of companion masses.
 
We sampled the parameter space on a 100$\times$100 period–mass grid, performing 10,000 integrations in total. Each system was integrated for 500 years $\sim$ 272,000 orbital periods of TOI-2109 b, sufficient to distinguish between regular and chaotic behavior over the explored parameter space \citep{2011IJNLM..46...23M}.

\begin{figure*}[ht]
    \centering
    \includegraphics[width=0.9\linewidth, height=10cm]{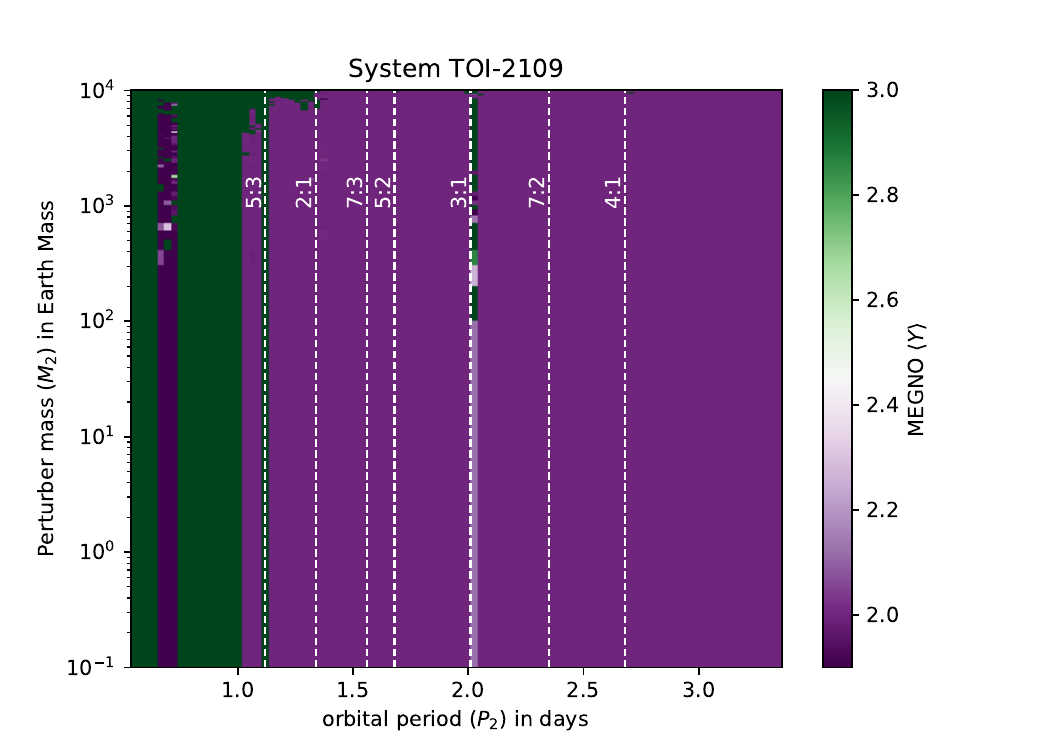}
    \caption{MEGNO map for the TOI-2109 two-planet system. Purple regions indicate dynamically stable configurations, while green regions indicate unstable configurations. White dotted lines mark the locations of mean-motion resonances.}
    \label{fig:megno_plot}
\end{figure*}

A companion was classified as ejected, and therefore dynamically unstable, if its distance from the host star exceeded 100 AU at any time during the integration, providing a practical upper limit for bound orbits. The resulting MEGNO map, shown in Figure~\ref{fig:megno_plot}, highlights regions of dynamical stability and the locations of prominent MMRs with TOI-2109 b, thereby constraining the parameter space for additional companions in the system.

\section{Identification of Mean Motion Resonances} \label{sec:DM}

The MEGNO map confirms that any viable perturbing companion must reside on an orbit exterior to that of TOI-2109 b, i.e., $P_2 > P_1$. 
The theoretical locations of mean motion resonances (MMRs) between TOI-2109 b and a hypothetical outer companion were computed from the standard period commensurability relation \citep{2012ApJ...761..122L}:

\begin{equation}
    \frac{P_2}{P_1} = \frac{j}{j - n},
\end{equation}

\noindent where $P_1$ and $P_2$ are the orbital periods of TOI-2109 b and the outer perturber, respectively, $j$ is a positive integer, and $n$ is the resonance order.
We evaluated all integer combinations with $j = 1$--$9$ and 
$n = 1$--$5$. The upper limit on resonance order was motivated by the low sinusoidal TTV amplitude recovered in our analysis $(A_{\mathrm{TTV}} \approx 44\, s)$ since higher-order resonances produce progressively weaker gravitational 
perturbations and correspondingly smaller observational effects \citep{Harre2024AA}. Where multiple $(j, n)$ combinations yield identical or near-identical period ratios, we retained only the lowest-order commensurability, as it dominates the dynamical interaction \citep{1999ssd..book.....M}. Resonances corresponding to 
$P_2 \leq 1.0$ \,day were excluded on dynamical grounds, as these fall within the chaotic inner zone identified by the MEGNO map. The 5:3 resonance was excluded, as the MEGNO map reveals predominantly 
chaotic behaviour at this commensurability and the 3:1 resonance  exhibiting stable orbits up to companion masses of $\approx$100\,$M_\oplus$. The MMR locations retained for further analysis are the 2:1, 3:1, 4:1, 5:2, 7:2, and 7:3 commensurabilities.
%we show that the preferred solution is not only statistically favoured but also dynamically viable.

\subsection{Differential Evolution}

Analytical formulations commonly used to estimate the mass of a 
perturbing companion from TTVs are generally valid only for first-order MMR's \citep{2012ApJ...761..122L, 2004astro.ph.12032A}. 
For higher-order resonances ($n > 1$), we therefore employed Differential Evolution (DE) \citep{1997JGOpt..11..341S}, a global optimization algorithm, to obtain robust initial estimates of the model parameters prior to the MCMC analysis. The optimization was performed using the \texttt{differential\_evolution} function implemented in the \texttt{scipy.optimize} module of the \texttt{SciPy} Python package. We adopted a population size of 15, a recombination probability of 0.7, a maximum of 1000 iterations, and a convergence tolerance of $10^{-4}$.
The DE algorithm minimizes the $\chi^{2}$ statistic between the observed mid-transit times and those predicted by the \texttt{TTVFast} dynamical model \citep{2014ApJ...787..132D}.
\texttt{TTVFast} is a fast and precise \(N\)-body integrator specifically designed for modeling transit timing variations in multi-planet systems. It requires six orbital elements as input for each planet:  orbital period ($P$), planetary mass ($M$), orbital inclination ($i$), eccentricity ($e$), argument of periastron ($\omega$), and longitude of the ascending node ($\Omega$).

Following convergence of the DE algorithm, the best-fitting solution was further refined using the L-BFGS-B local optimization method, which efficiently locates the nearest local minimum in parameter space by minimizing the same $\chi^{2}$ function within the specified parameter bounds.
To reduce the possibility of convergence to local minima and to examine the consistency of the optimization results, the DE procedure was independently repeated approximately ten times for each investigated resonance configuration. Since DE is a stochastic, population-based optimization algorithm, separate runs may converge to slightly different regions of the parameter space. For each resonance configuration, we selected the solution with the lowest $\chi_\nu^{2}$, obtained across all optimization runs. This best-fitting parameter set was then used to initialize the subsequent MCMC analysis. The parameter bounds adopted for the DE search and their best values are listed in Table~\ref{table:7:3_DEvalues}.

\begin{table}
    \centering
    \caption{Initial parameter guesses for the Differential Evolution (DE) optimization.}
    \label{table:7:3_DEvalues}
    
    \begin{tabular}{ccc}
        \hline
        \hline
        Parameter & Priors (Uniform) & Best-fit values \\
        \hline
        \hline
        $m_1\ (M_{\oplus})$ & [1361, 1829] & 1632.37 \\
        $m_2\ (M_{\oplus})$ & [1.0, 700] & 218.71 \\
        $P_1\ (\mathrm{days})$ & [0.67237, 0.67257] & 0.67248 \\
        $P_2^{\dagger}\ (\mathrm{days})$ & [1.556,1.589] & 1.578 \\
        $M_{a,1}\ (\mathrm{deg})$ & [0, 360] & 0.0007 \\
        $M_{a,2}\ (\mathrm{deg})$ & [0, 360] & 189.06 \\
        \hline
        
    \end{tabular}
    
     \footnotesize
    $^{\dagger}$ The prior range for $P_2$ was adjusted according to the mean-motion resonance considered. The value shown here corresponds to the 7:3 resonance.
\end{table}

\subsection{Dynamical modeling of TTVs} 

To investigate whether the observed transit timing variations (TTVs) of TOI-2109 b can be explained by gravitational perturbations from an additional planetary companion, and to characterize its physical and orbital properties, we employed \texttt{TTVFast}.
%\citep{2014ApJ...787..132D}.TTVFast is a fast and precise \(N\)-body integrator specifically designed for modeling transit timing variations in multi-planet systems.
%\texttt{TTVFast} requires six orbital elements as input for each planet:  orbital period ($P$), planetary mass ($M$), orbital inclination ($i$), eccentricity ($e$), argument of periastron ($\omega$), and longitude of the ascending node ($\Omega$). 
The available transit-timing measurements do not provide sufficient information to independently constrain all the parameters of the two-planet model.  Moreover, generating dynamical TTV models across the full observational baseline of approximately 20 years (2006-2025) is computationally expensive. We therefore adopted a simplified, physically motivated orbital configuration to reduce the dimensionality of the parameter space:

\begin{enumerate}
\item  Nodal Angles: The longitudes of the ascending node were fixed to $\Omega_{b}=\Omega_{c}=180^{\circ}$. 

\item Coplanar: The system was assumed to be coplanar, such that $i_{b}=i_{c}$.

\item Circular orbits: Both planets were assumed to have circular orbits ($e_b = e_c = 0$). For TOI-2109\,b, this is consistent with tidal circularization expected at such an extreme orbital separation. 
Consequently, the arguments of periastron, $\omega_{b}$ and $\omega_{c}$, are undefined and were not included as free parameters.
\end{enumerate}
This simplified model provides a practical and physically reasonable approach for exploring the TTV signal.

\begin{figure*}
    \centering
    \includegraphics[width=0.9\linewidth, height=10cm]{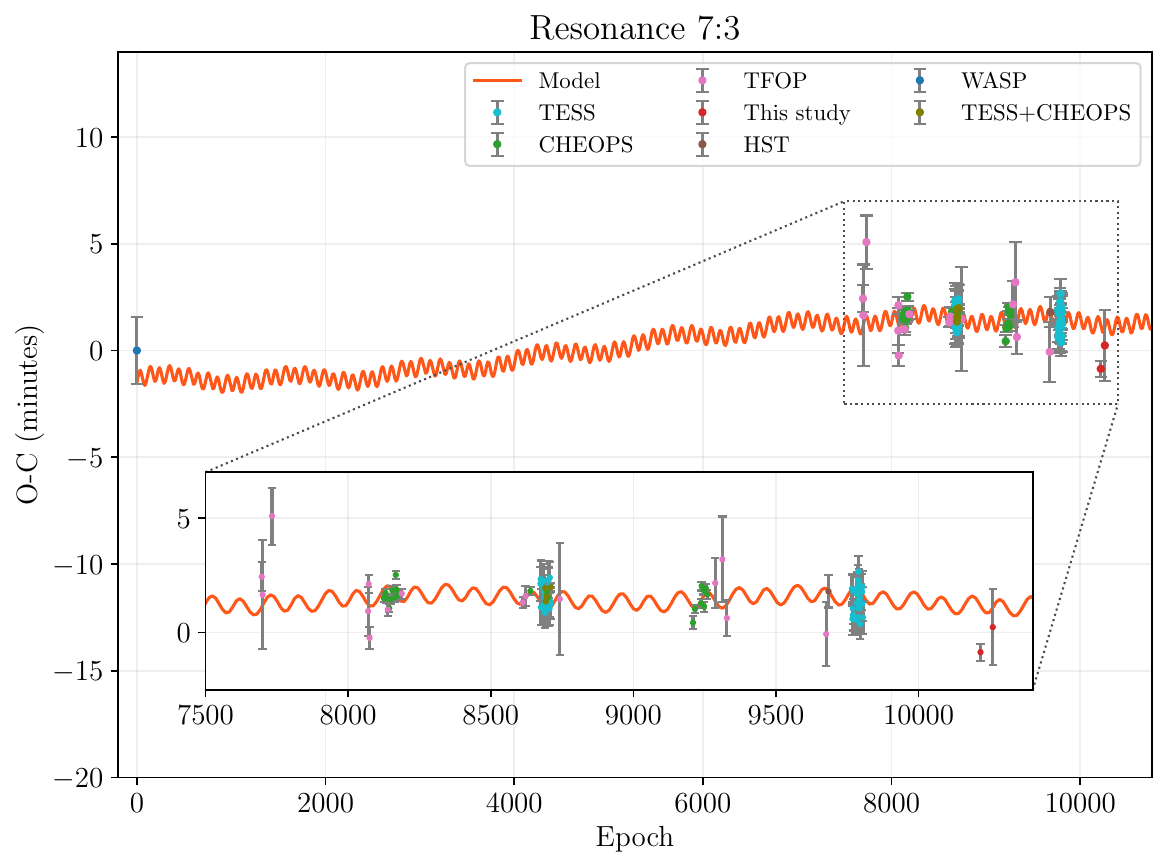}
    \caption{Observed transit timing variations of TOI-2109b compared to the best-fit two-planet dynamical model. The solid line shows the predicted TTV signal for the 7:3 mean-motion resonance.}
    \label{fig:twoplanet_model}
\end{figure*}

To sample the posterior distributions of the model parameters, \texttt{TTVFast}  was coupled with an MCMC sampler following the same procedure described in 
Section~\ref{sec:ttvanalysis}. The best-fitting solutions obtained from the DE  optimization were used as initial guesses for the MCMC chains, which were run 
for 60,000 steps with a burn-in of 10,000 steps and 50 walkers. The resulting  TTV model and corresponding corner plot are shown in  Figures~\ref{fig:twoplanet_model} and \ref{fig:cornor_ttvfast}, respectively.

The dynamical TTV model was computed over the full observational baseline of 20 years (2006-2025), covering 87 predicted transit times, which 
represents one of the longest baselines modeled dynamically for a system of this type. 

Among all resonance configurations tested, the 7:3 MMR provides 
the best fit to the observed TTVs, as quantified by the lowest $\chi^2_{\nu} =1.83$ and lowest BIC  value. This result is consistent with \citet{Harre2024AA} and suggests that the additional planet is on a circular orbit in a 7:3 MMR. The best-fitting orbital and physical parameters for all tested resonances are summarized in Table~\ref{tab:7:3_values}.

\begin{table*}
    \centering
    \caption{Best-fit MCMC results for the two-planet model.}
    \label{tab:7:3_values}
   
    \begin{tabular}{cccc}
        \hline
        \hline
        Parameter & Priors & Distribution & Best-fit values \\
        \hline
        \hline
        $m_1\ (M_{\oplus})$ & $[1604,1652]$ & Uniform & $1634.61 \pm 9.85$ \\
        $m_2\ (M_{\oplus})$ & $[203,239]$ & Uniform & $222.20 \pm 6.44$ \\
        $P_1\ (\mathrm{days})$ & $(0.6724,\ 2.8\times10^{-7})$ & Gaussian & $0.6724894 \pm 2.65\times10^{-7}$ \\
        $P_2\ (\mathrm{days})$ & $(1.578,0.001)$ &Gaussian & $1.57813\pm+8.9\times 10^{-5} $\\
        $M_{a,1}\ (\mathrm{deg})$ & $(0.0007,\ 1)$ & Gaussian & $0.064 \pm 0.718$ \\
        $M_{a,2}\ (\mathrm{deg})$ & $(189.24,\ 1)$ & Gaussian & $189.40 \pm 0.64$ \\
        \hline
        
    \end{tabular}
\end{table*}

% twoplanet model

\subsection{Model Selection}
To identify the model that best describes the observed TTVs of TOI-2109 b, we compared five commonly used timing models using $\chi_\nu^2$ and the Bayesian Information Criterion (BIC). The BIC is defined as
\begin{equation}
\mathrm{BIC} = \chi^2 + k \ln n,
\end{equation}

where  $k$ is the number of free parameters and $n$ is the number of data points. The values of $\mathbf{\chi^2}$, $\chi_\nu^2$, number of free parameters, and BIC values for all tested models are listed in Table~\ref{table:5models}.

We first compared three baseline timing models: a linear ephemeris, an orbital decay model, and an apsidal precession model. The apsidal precession model provides the best fit among them, with $\Delta\mathrm{BIC}_{\mathrm{linear-apsidal}} = 6.37$ and $\Delta\mathrm{BIC}_{\mathrm{decay-apsidal}} = 1.24.$ 
However, none of these models fully describe the structure of the TTVs. The sinusoidal model is preferred over the apsidal precession model, with $\Delta\mathrm{BIC} = 20.6$, which shows very strong evidence for a periodic signal in the timing residuals. Given this evidence for periodicity, we tested a two-planet dynamical model.
We examined a range of plausible MMRs, among these, the 7:3 resonance provides the best fit. Relative to the other resonances, 
the 7:3 model is strongly preferred, with $\Delta\mathrm{BIC}_{2:1-7:3} = 39.57$, $\Delta\mathrm{BIC}_{3:1-7:3} = 70.27$, $\Delta\mathrm{BIC}_{4:1-7:3} = 26.37$, $\Delta\mathrm{BIC}_{5:2-7:3} = 48.80$, and $\Delta\mathrm{BIC}_{7:2-7:3} = 16.79$.  Compared with the sinusoidal model, the two-planet model is preferred, with $\Delta\mathrm{BIC} = 4.95$, indicating positive evidence in its favor. We therefore adopt the two-planet dynamical model as the most plausible explanation for the observed TTVs in the TOI-2109~b system.

%of $\chi_\nu^2 = 2.342$, $2.264$, and $2.182$, respectively. the clear residual structure in the $O\!-\!C$ diagram (see Fig.~\ref{fig:ocplot_TOI-2109b}) indicates that a sinusoidal variations.

%timing models
\begin{table*}
\centering
\caption{Timing models tested for TOI-2109b.}
\label{table:5models}
\begin{tabular}{lcccc}
\hline\hline
Model & No.of Parameters &$\chi^2$ &$\chi_\nu^2$ & BIC \\
\hline\hline
Linear model  & 3 &196.85 &2.34 & 210.25 \\
Orbital decay model & 4 & 187.26&2.25& 205.12\\
Apsidal precession model & 6 & 177.08&2.18 & 203.88 \\
Sinusoidal model  & 4 &165.53 &1.99 & 183.21\\
\cutinhead{Two-planet model}
2:1 & 7&186.58 &2.33&217.83\\
3:1 & 7&217.27 &2.71&248.53\\
4:1 & 7& 173.37&2.16&204.63\\
5:2 & 7& 195.80&2.44&227.06\\
7:2 & 7& 163.79&2.04&195.05\\
7:3 & 7 & 147.08& 1.83 &178.26 \\
\hline
\end{tabular}
\end{table*}

\section{The Concluding Remarks} \label{sec:R&D}
We have presented a comprehensive transit timing analysis of the UHJ TOI-2109\,b using observations obtained from WASP, TESS, CHEOPS, HST, TFOP, and new ground-based photometry. This represents the longest transit-timing baseline (2006-2025) currently available for a UHJs, substantially improving sensitivity to long-term dynamical signals compared with previous studies. 

We tested five models representing distinct physical origins for the observed timing variations. The linear model indicates that the orbital period is not strictly constant.
The decay fit gives $\dot{P} = -3.96 \pm 2.05 $~ms\,yr$^{-1}$, corresponding to a stellar tidal quality factor of $Q_\star' \sim 1.85\times10^{7}$, consistent with the weak tidal dissipation expected for an F-type host star \citep{2020MNRAS.498.2270B}. 

The apsidal precession model provides the best statistical fit, with $\Delta\mathrm{BIC}_{\rm linear-apsidal} = 6.37$ and $\Delta\mathrm{BIC}_{\rm decay-apsidal} = 1.24$, it fails to fully explain the O-C residuals. A GLS periodogram reveals a periodic signal with a FAP of 2.34\% and a TTV amplitude of $\approx 44$~s. A sinusoidal model improves both $\chi_\nu^2$ and the BIC relative to the apsidal model ($\Delta\mathrm{BIC} = 20.6$), suggesting that the observed TTVs contain a periodic signal rather than random timing noise.
Motivated by this periodicity, we performed $N$-body dynamical modelling with \texttt{TTVFast}. MEGNO stability maps identified several dynamically viable MMRs, and among these, the 7:3 MMR consistently yielded the lowest $\chi_\nu^2$ following differential-evolution optimization and subsequent MCMC refinement. The best-fitting solution corresponds to a non-transiting outer companion with an orbital period of $1.57813 \pm 8.9\times10^{-5}$ days and a mass of $222.20 \pm 6.44\,M_\oplus$, which reproduces the observed timing variations substantially better than any analytical model. 

The two-planet model yields both the lowest $\chi_\nu^2$ (1.83) and the lowest BIC (178.20) of all models tested. Relative to the sinusoidal model, it is favored by $\Delta\mathrm{BIC} = 4.95$, providing positive evidence for TTVs driven by gravitational perturbations from an additional, non-transiting companion near the 7:3 mean-motion resonance.

This result is broadly consistent with \citet{Harre2024AA}, who proposed a nearby companion with a period close to 1.57 days in a circular 7:3 resonant orbit based on a lower-significance signal. Our longer timing baseline, comparison across five models, and self-consistent 
$N$-body fit strengthen this scenario, giving a lower $\chi_\nu^2$ and a more tightly constrained companion solution than previous work.

Stellar activity is an unlikely origin for the observed TTVs. \citet{2021AJ....162..256W} used TESS photometry to show that the transit timing variations of TOI-2109 b do not correlate with stellar activity indicators. This is consistent with the coherent, periodic nature of the TTVs found in our analysis.

Although the dynamical model is statistically favoured, several limitations should be noted. The TTVFast analysis assumes coplanar and circular orbits to reduce the dimensionality of the parameter space and maintain computational feasibility over the nearly two-decade observational baseline.  While these assumptions are physically motivated for an UHJs, future timing data with higher precision may allow more general eccentric or mutually inclined configurations to be explored.
Confirming the planetary nature of the proposed companion will require continued transit monitoring, which will help refine its orbital parameters and further test the stability of the resonant configuration.
\clearpage
\begin{acknowledgments}
\section*{Acknowledgments}
We are grateful to the anonymous referee for 
good comments and suggestions. 
We acknowledge funding from the National Science and Technology Council (NSTC), Taiwan, through Grants NSTC 113-2112-M-007-030, NSTC 114-2112-M-007-029.
% LCOGT
This work makes use of observations from the LCOGT network. Part of the LCOGT telescope time was granted by NOIRLab through the Mid-Scale Innovations Program (MSIP). MSIP is funded by NSF.
% LCOGT 0m35 Delta Rho observations  (new 0m4 / 0m35)
This paper is based on observations made with the Las Cumbres Observatory’s education network telescopes that were upgraded through generous support from the Gordon and Betty Moore Foundation.
% ExoFOP
This research has made use of the Exoplanet Follow-up Observation Program (ExoFOP; DOI: 10.26134/ExoFOP5) website, which is operated by the California Institute of Technology, under contract with the National Aeronautics and Space Administration under the Exoplanet Exploration Program
% TESS
Funding for the TESS mission is provided by NASA's Science Mission Directorate. KAC acknowledges support from the TESS mission via subaward s3449 from MIT and NASA grants 80NSSC24K1889 and 80NSSC26K0081
% MuSCAT3
This is paper is based on observations made with the MuSCAT3 instrument, developed by the Astrobiology Center and under financial supports by JSPS KAKENHI (JP18H05439) and JST PRESTO (JPMJPR1775), at Faulkes Telescope North on Maui, HI, operated by the Las Cumbres Observatory.
This work made use of data from the TESS, CHEOPS, and HST/WFC3 missions, 
as well as ground-based follow-up photometry of TESS targets obtained through the \dataset[TFOP]{https://exofop.ipac.caltech.edu/tess/} program. TESS data are publicly available through the 
\dataset[Mikulski Archive for Space Telescopes]{https://mast.stsci.edu/portal/Mashup/Clients/Mast/Portal.html} (MAST) portal from\dataset[https://doi.org/10.17909/t9-nmc8-f686]{https://doi.org/10.17909/t9-nmc8-f686}. 
CHEOPS data analyzed in this paper are available in the 
\dataset[CHEOPS]{https://cheops-archive.ssdc.asi.it/archive_browser/} mission archive. The publicly available HST observations presented here were taken as part of proposal 17543, led by Guangwei Fu, and were obtained through the MAST archive from 
\dataset[https://doi.org/10.17909/m1vw-1z37]{ https://doi.org/10.17909/m1vw-1z37}. Two nights of observational data were obtained using the 1-meter telescope at the Thai National Observatory, operated by the National Astronomical Research Institute of Thailand (NARIT). We gratefully acknowledge the TNO staff for their support during the observations.

\end{acknowledgments}

\facilities{Thai National Observatory(TNO)-1\,m}

\software{emcee \citep{2013PASP..125..306F}, PyLightcurve \citep{2016ApJ...832..202T}, REBOUND \citep{2016MNRAS.459.2275R}, TTVFast \citep{2014ApJ...787..132D}, Lightkurve \citep{2018ascl.soft12013L}, AstroImageJ \citep{2017AJ....153...77C}, SciPy \citep{virtanen2020scipy}, PyAstronomy \citep{2019ascl.soft06010C}, Iraclis \citep{2016ApJ...832..202T}.}

\appendix

\renewcommand{\thefigure}{\thesection\arabic{figure}}
\renewcommand{\thetable}{\thesection\arabic{table}}

%------------------------------------------------
\section{CHEOPS and HST Individual light curves}
\label{app:lightcurves}
\setcounter{figure}{0}

% CHEOPS
\begin{figure}[ht]
    \centering
    \includegraphics[width=0.4\linewidth, height=18cm]{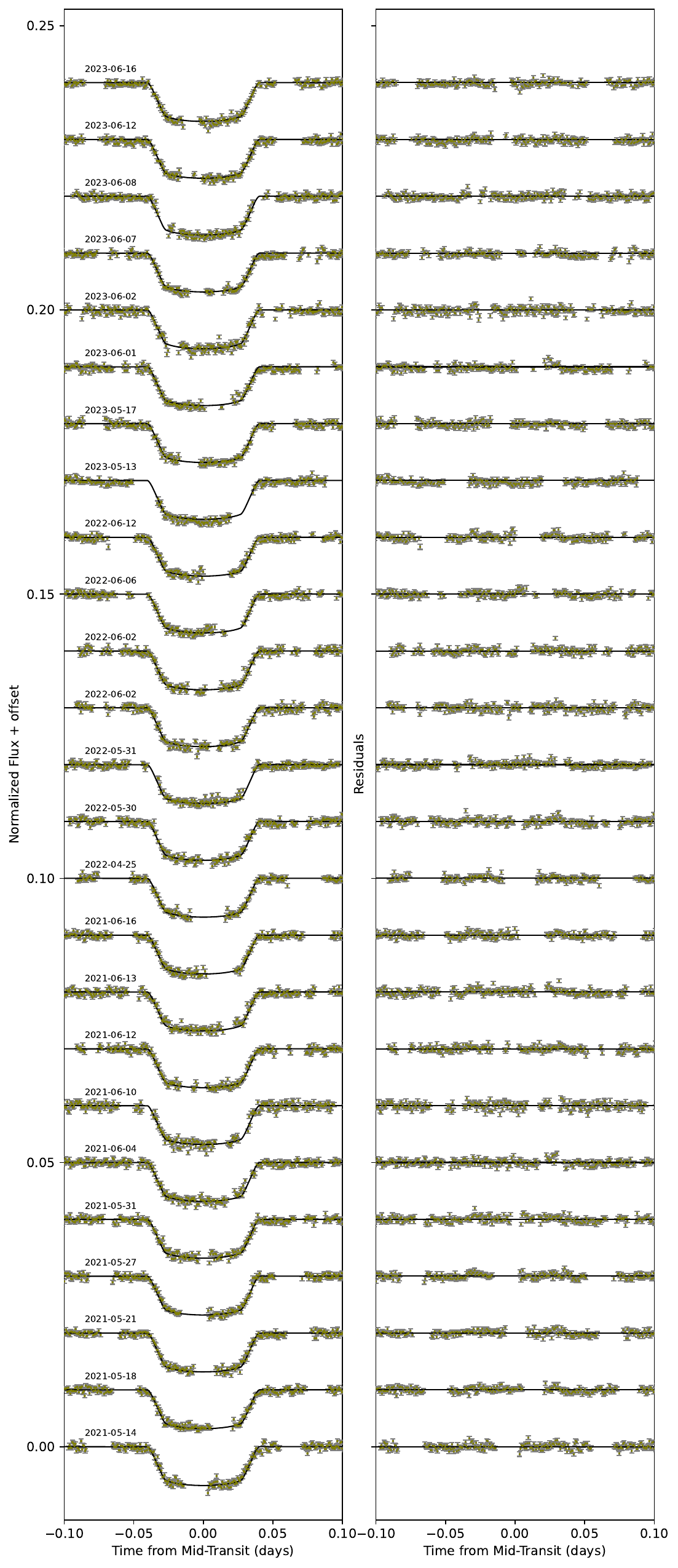}
    \caption{Individual CHEOPS light curves of TOI-2109b, shown with the best-fit transit model (solid line) and its residuals.}
    \label{app:cheops_25}
\end{figure}

% HST
\begin{figure}[ht]
    \centering
    \includegraphics[width=0.4\linewidth, height=20cm]{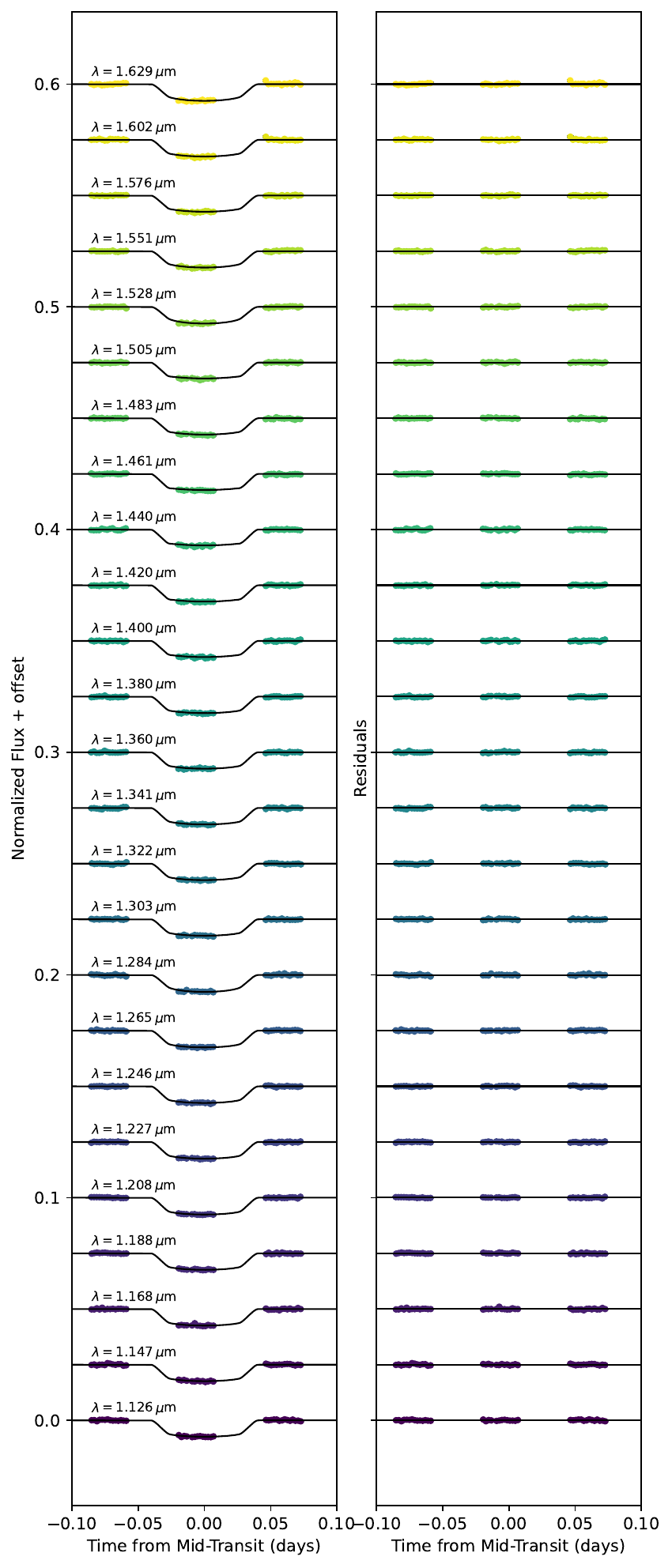}
    \caption{HST spectroscopic light curves of TOI-2109b in 25 wavelength bins shown with the best-fit transit model (solid line) and its residuals.}
    \label{app:25_hst}
\end{figure}
%keepaspectratio
\clearpage
%------------------------------------------------
\section{Posterior Probability Distribution for the five Models used in our work}
\label{app:corner}
\setcounter{figure}{0}

% Linear
\begin{figure*}[ht]
    \centering
    \includegraphics[width=0.6\textwidth, height=10cm, keepaspectratio]{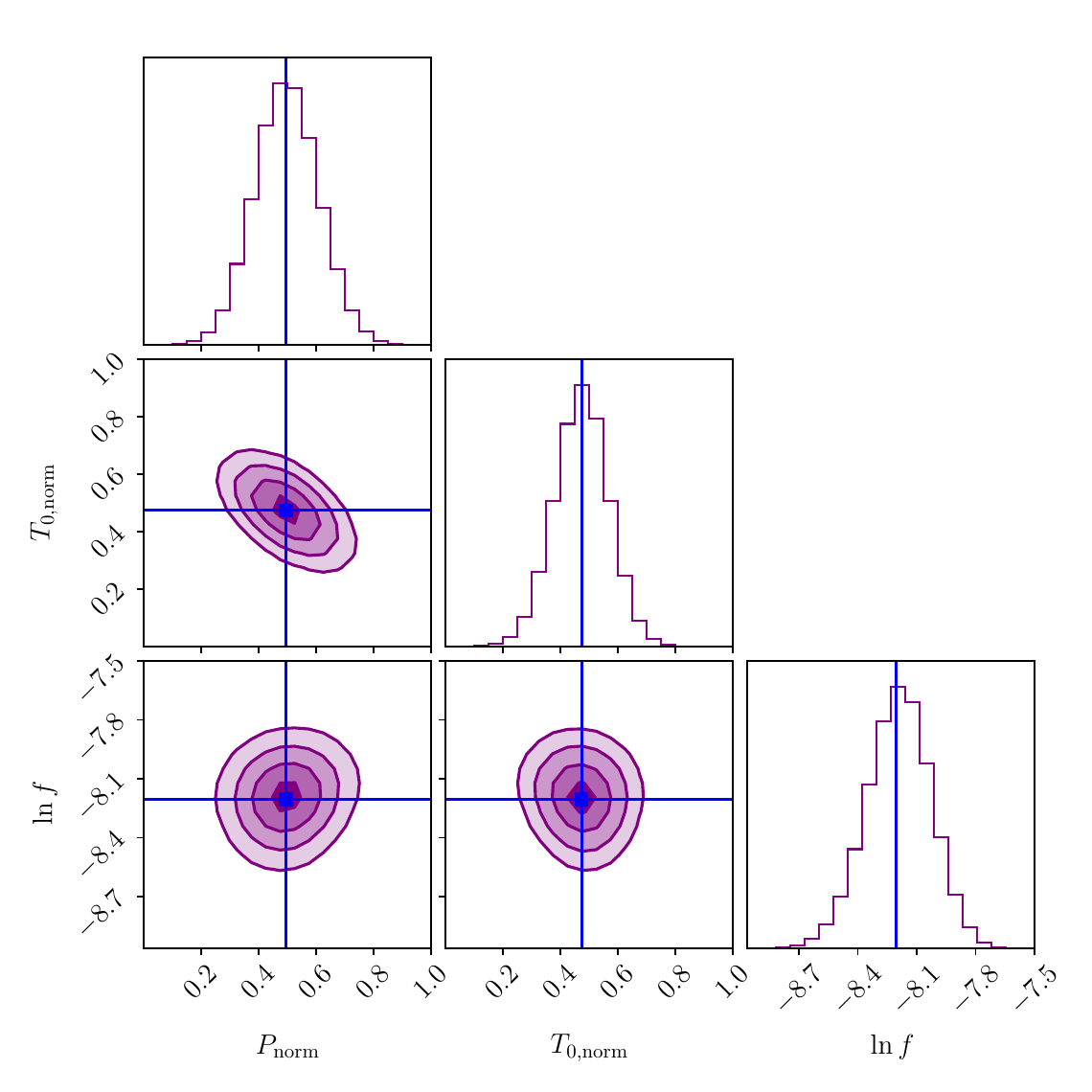}
    \caption{Corner plot of posterior distributions from the linear model MCMC fit for TOI-2109 b. $P_{\mathrm{norm}}$ and $T_{0,\mathrm{norm}}$ are normalized to the range $[0, 1]$ for visualization. The blue lines indicate the best-fit MAP values.}
    \label{fig:Linearmodel_cornor}
\end{figure*}

% Orbital decay
\begin{figure*}[ht]
    \centering
    \includegraphics[width=0.6\textwidth, height=10cm, keepaspectratio]{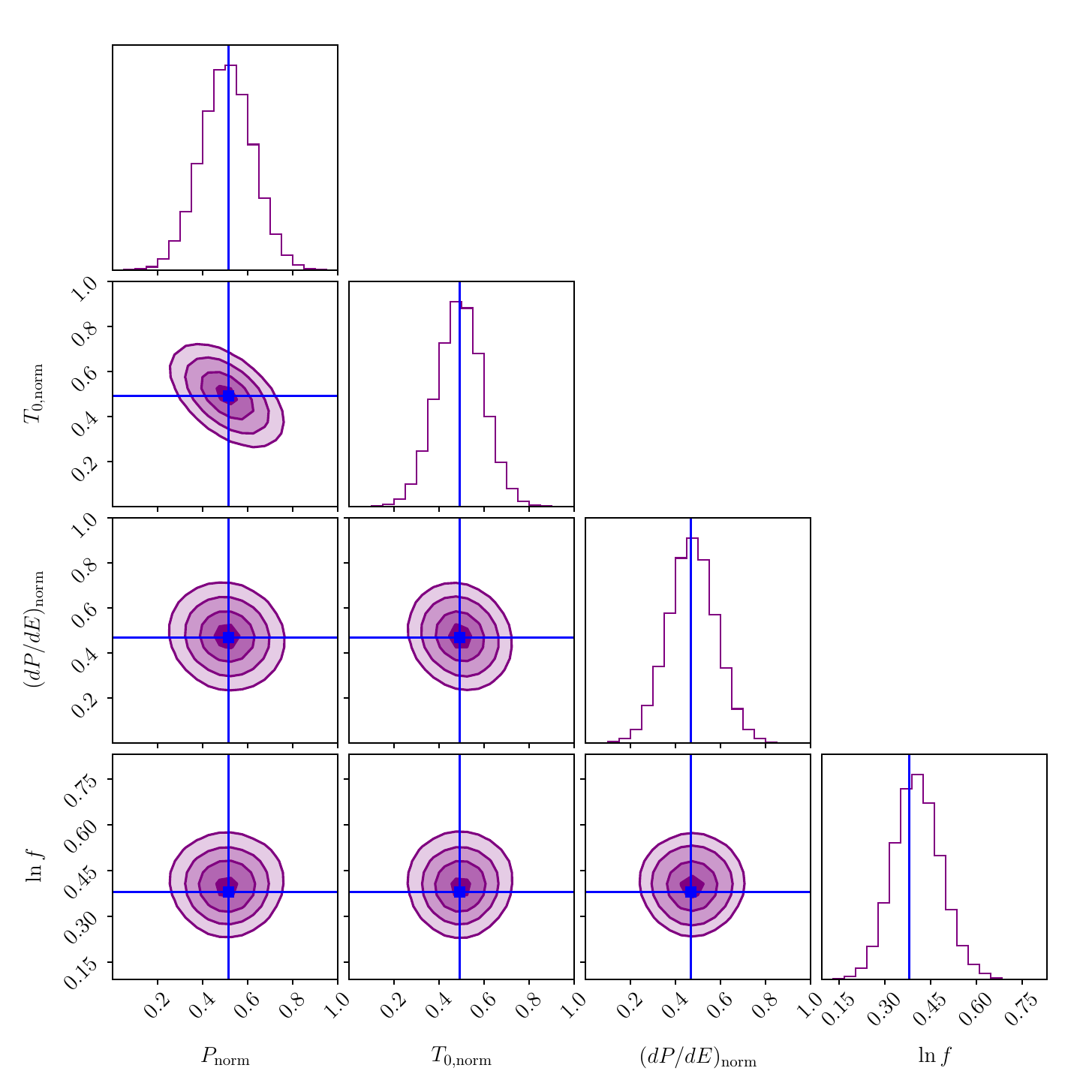}
    \caption{Corner plot of posterior distributions from the orbital decay model MCMC fit for TOI-2109 b. $P_{\mathrm{norm}}$, $T_{0,\mathrm{norm}}$ and $(dP/dE)_{\mathrm{norm}}$ are normalized to the range $[0, 1]$ for visualization. The blue lines indicate the best-fit MAP values.}
    \label{fig:decaymodel_cornor}
\end{figure*}

% Apsidal precession
\begin{figure*}[ht]
    \centering
    \includegraphics[width=0.8\textwidth, height=14cm, keepaspectratio]{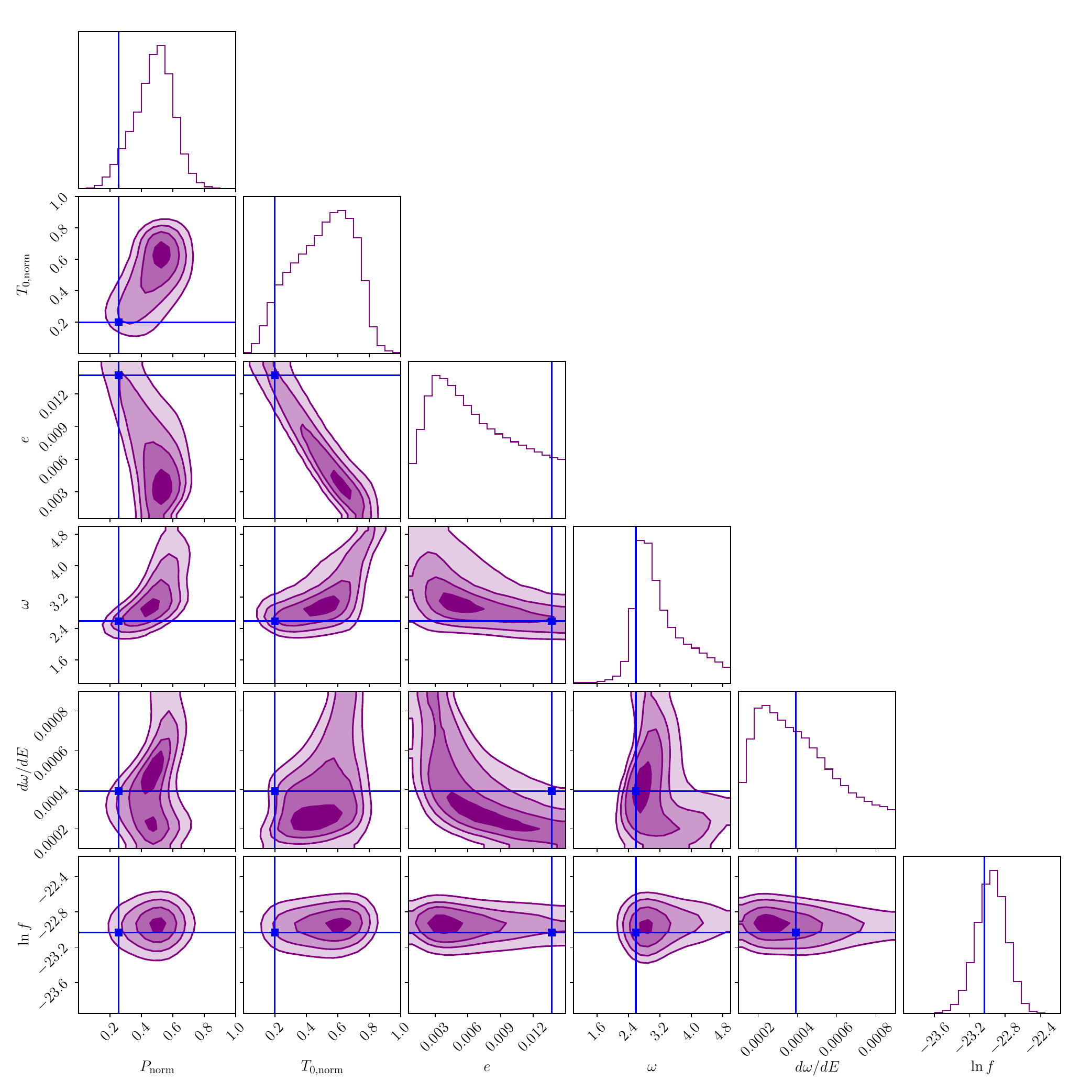}
    \caption{Corner plot of posterior distributions from the apsidal precession model MCMC fit for TOI-2109 b. $P_{\mathrm{norm}}$ and $T_{0,\mathrm{norm}}$ are normalized to the range $[0, 1]$ for visualization. The blue lines indicate the best-fit MAP values.}
    \label{fig:Apsidalmodel_cornor}
\end{figure*}

% Sinusoidal
\begin{figure*}[ht]
    \centering
    \includegraphics[width=0.9\textwidth, height=14cm, keepaspectratio]{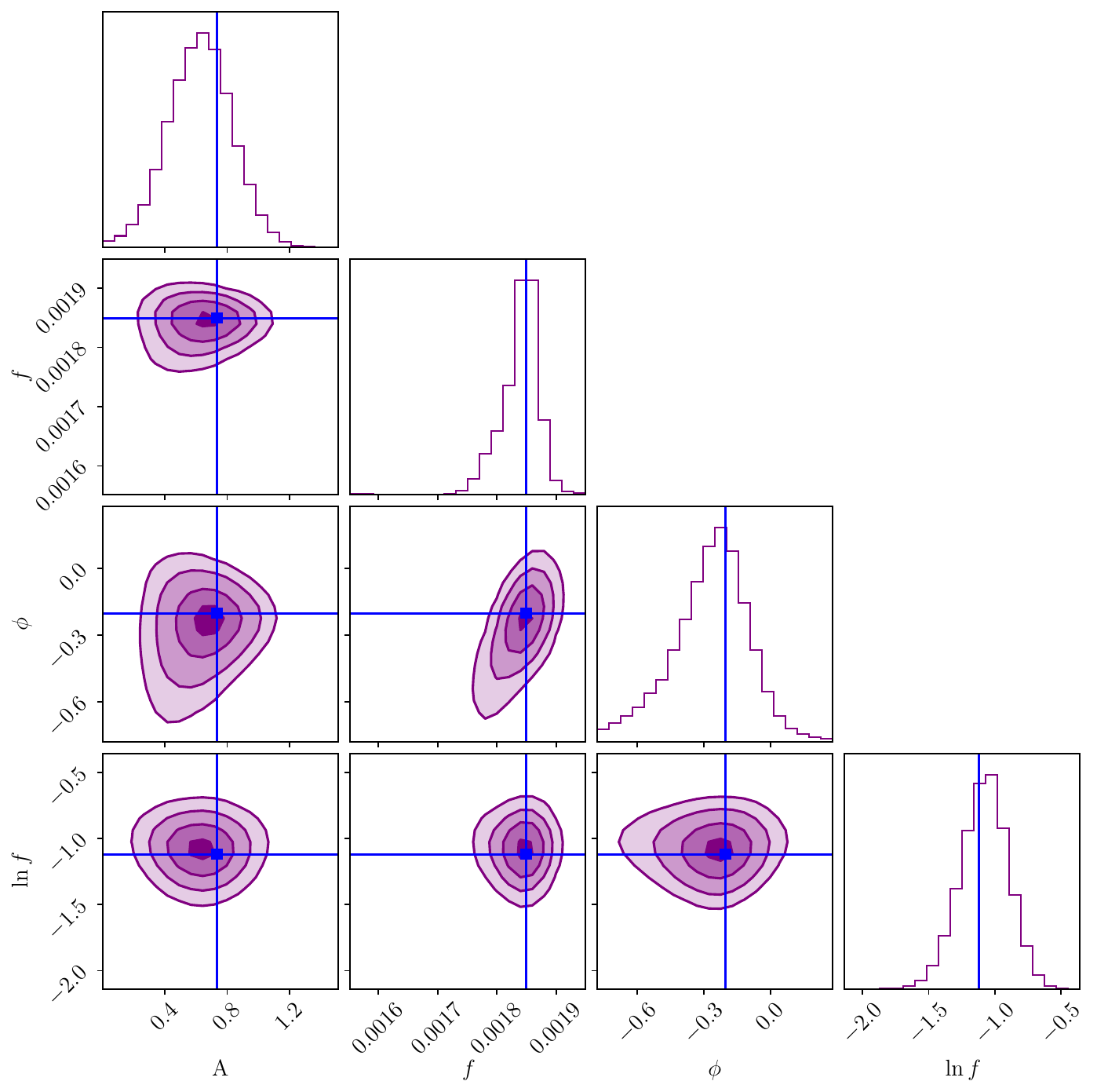}
    \caption{Corner plot of posterior distributions from the sinusoidal model MCMC fit for TOI-2109 b. The blue lines indicate the best-fit MAP values.}
    \label{fig:sinusoidal_cornor}
\end{figure*}

% Two-planet
\begin{figure*}[ht]
    \centering
    \includegraphics[width=0.9\textwidth, height=16cm, keepaspectratio]{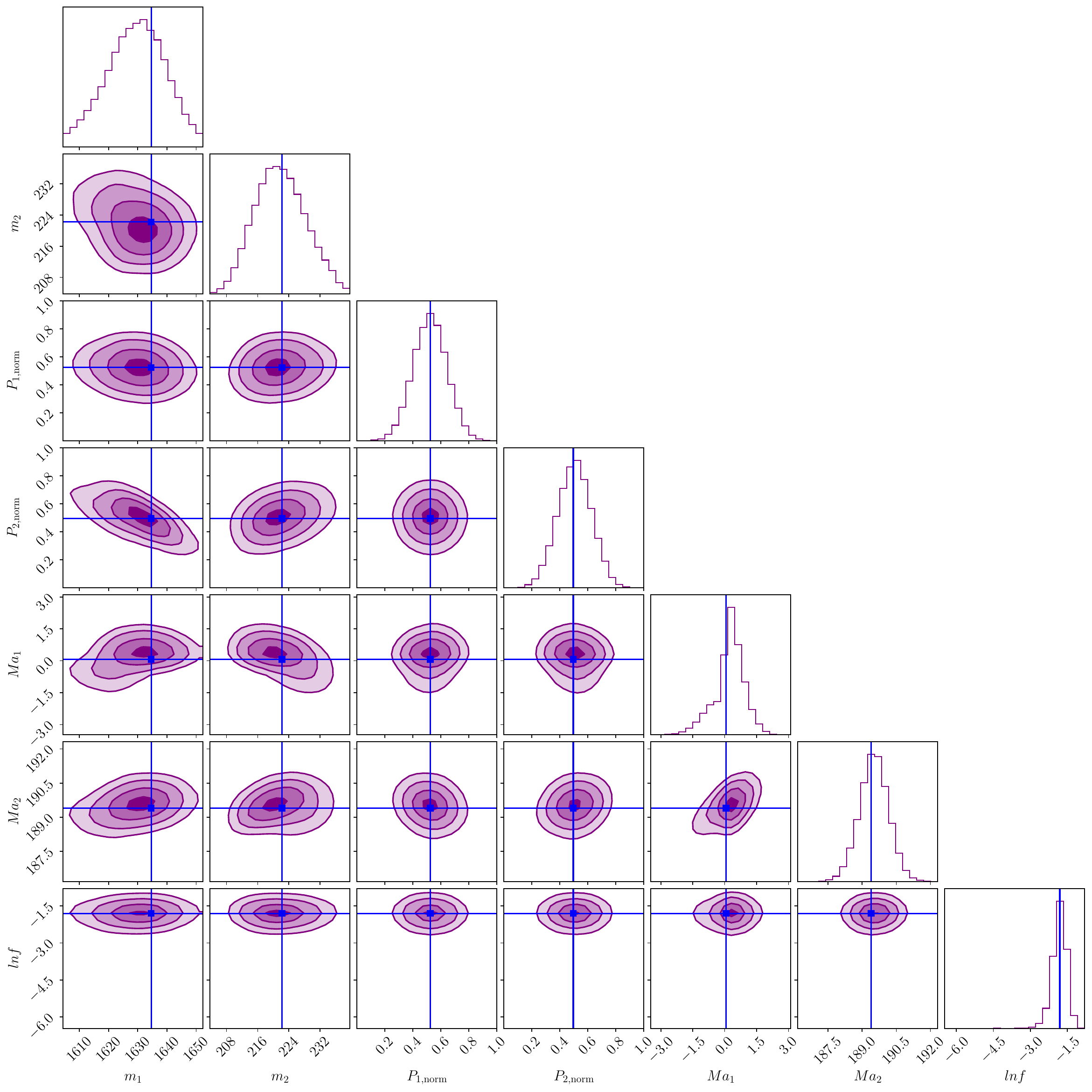}
    \caption{Corner plot of posterior distributions from the two-planet model MCMC fit for TOI-2109 b. $P_{1,\mathrm{norm}}$ and $P_{2,\mathrm{norm}}$ are normalized to the range $[0, 1]$ for visualization. The blue lines indicate the best-fit MAP values.}
    \label{fig:cornor_ttvfast}
\end{figure*}
\clearpage
%------------------------------------------------
\section{Tables}
\label{app:midtimes}
\setcounter{table}{0}

%CHEOPS observation log table
\startlongtable
\begin{deluxetable*}{cccc}
\tablecaption{The CHEOPS observational log of TOI-2109 b. \label{tab:cheops_log}}
\tablehead{
\colhead{Visit} & \colhead{Date (UTC)} & \colhead{Duration (hr)} & \colhead{Aperture (pixels)}
}
\startdata
1  & 2021-05-14 22:40:47 & 5.29  & 18\\
2  & 2021-05-18 23:05:53 & 5.62  & 25\\
3  & 2021-05-21 16:00:37 & 5.44  & 19\\
4  & 2021-05-27 17:05:41 & 5.42  & 22\\
5  & 2021-05-31 01:46:02 & 5.64  & 20\\
6  & 2021-06-04 02:41:41 & 5.64  & 20\\
7  & 2021-06-10 19:49:56 & 5.64  & 15\\
8  & 2021-06-12 04:41:34 & 5.95  & 21\\
9  & 2021-06-13 12:38:51 & 5.60  & 40\\
10 & 2021-06-16 04:55:27 & 5.64  & 20\\
11 & 2022-04-10 19:24:47 & 5.54  & 39$^{\dagger}$\\
12 & 2022-04-16 04:06:53 & 5.64  & 39$^{\dagger}$\\
13 & 2022-04-25 14:28:53 & 6.44  & 24\\
14 & 2022-05-30 13:35:04 & 5.59  & 18\\
15 & 2022-05-31 05:37:44 & 5.64  & 23$^{\dagger}$\\
16 & 2022-05-31 21:47:04 & 5.69  &23 \\
17 & 2022-06-02 22:24:03 & 5.59  & 17\\
18 & 2022-06-02 06:18:44 & 5.45  & 22\\
19 & 2022-06-06 06:43:02 & 5.64  & 19\\
20 & 2022-06-12 07:58:36 & 5.64  & 21\\
21 & 2023-04-15 00:13:48 & 5.62  & 40$^{\dagger}$\\
22 & 2023-04-21 01:33:16 & 5.34  & 15$^{\dagger}$\\
23 & 2023-04-27 02:41:40 & 5.25  & 16$^{\dagger}$\\
24 & 2023-05-13 07:55:27 & 23.74 & 19\\
25 & 2023-05-17 21:47:15 & 26.00 & 19\\
26 & 2023-06-01 06:32:26 & 25.35 & 23\\
27 & 2023-06-02 08:28:44 & 23.48 & 17\\
28 & 2023-06-07 09:18:43 & 24.68 & 22\\
29 & 2023-06-08 19:05:01 & 23.56 & 21\\
30 & 2023-06-12 19:20:57 & 23.74 & 19\\
31 & 2023-06-16 03:34:32 & 23.74 & 23\\
\enddata
\tablecomments{$^{\dagger}$ These light curves were excluded from the analysis.}
\end{deluxetable*}

%midtransit table
\startlongtable
\begin{deluxetable*}{cccc}
\tablecaption{Mid-transit times of TOI-2109 b used in this analysis. \label{app:midtimes_all}}
\tablehead{
\colhead{Epoch} & \colhead{Mid-transit time (BJD$_{\rm TDB}$)} & \colhead{O$-$C (days)} & \colhead{Source}
}
\startdata
-8168 & $2453885.688618 \pm 0.001079$ & -0.002060 & (a) \\
-471 & $2459061.724320 \pm 0.000440$ & 0.000510 & (b) \\
-468 & $2459063.741200 \pm 0.001650$ & -0.000032 & (b) \\
-435 & $2459085.935230 \pm 0.000865$ & 0.002352 & (b) \\
-98 & $2459312.556160 \pm 0.000740$ & -0.000489 & (b) \\
-96 & $2459313.901930 \pm 0.000270$ & 0.000333 & (b) \\
-93 & $2459315.917720 \pm 0.000345$ & -0.001299 & (b) \\
-43 & $2459349.542634 \pm 0.000125$ & -0.000090 & (c) \\
-37 & $2459353.577640 \pm 0.000124$ & 0.000071 & (c) \\
-33 & $2459356.267370 \pm 0.000135$ & -0.000095 & (c) \\
-29 & $2459358.956930 \pm 0.000185$ & -0.000431 & (b) \\
-24 & $2459362.319340 \pm 0.000125$ & -0.000392 & (c) \\
-19 & $2459365.681980 \pm 0.000130$ & -0.000122 & (c) \\
-13 & $2459369.717100 \pm 0.000140$ & 0.000153 & (c) \\
-3 & $2459376.441640 \pm 0.000155$ & -0.000048 & (c) \\
-1 & $2459377.787260 \pm 0.000130$ & 0.000624 & (c) \\
1 & $2459379.131760 \pm 0.000130$ & 0.000176 & (c) \\
5 & $2459381.821480 \pm 0.000140$ & -0.000001 & (c) \\
20 & $2459391.908650 \pm 0.000150$ & 0.000058 & (b) \\
444 & $2459677.037450 \pm 0.000160$ & -0.000160 & (b) \\
454 & $2459683.762380 \pm 0.000300$ & 0.000029 & (b) \\
471 & $2459695.194600 \pm 0.000125$ & 0.000190 & (c) \\
506 & $2459718.731420 \pm 0.000515$ & 0.000416 & (d) \\
507 & $2459719.404030 \pm 0.000565$ & 0.000552 & (d) \\
508 & $2459720.075650 \pm 0.000545$ & -0.000302 & (d) \\
516 & $2459725.456020 \pm 0.000600$ & 0.000275 & (d) \\
517 & $2459726.128720 \pm 0.000520$ & 0.000501 & (d) \\
518 & $2459726.800340 \pm 0.000480$ & -0.000353 & (d) \\
519 & $2459727.473340 \pm 0.000500$ & 0.000173 & (d) \\
520 & $2459728.145450 \pm 0.000490$ & -0.000191 & (d) \\
521 & $2459728.818120 \pm 0.000515$ & 0.000005 & (d) \\
522 & $2459729.490120 \pm 0.000470$ & -0.000469 & (d) \\
523 & $2459730.163341 \pm 0.000125$ & 0.000278 & (c),(d)$^{\dagger}$ \\
525 & $2459731.507889 \pm 0.000193$ & -0.000123 & (c),(d)$^{\dagger}$ \\
526 & $2459732.180150 \pm 0.000540$ & -0.000336 & (d) \\
527 & $2459732.852980 \pm 0.000127$ & 0.000020 & (c),(d)$^{\dagger}$ \\
528 & $2459733.525600 \pm 0.000140$ & 0.000166 & (c) \\
533 & $2459736.887810 \pm 0.000130$ & 0.000006 & (c) \\
534 & $2459737.560620 \pm 0.000575$ & 0.000342 & (d) \\
535 & $2459738.233290 \pm 0.000565$ & 0.000537 & (d) \\
536 & $2459738.905540 \pm 0.000575$ & 0.000313 & (d) \\
537 & $2459739.577430 \pm 0.000590$ & -0.000271 & (d) \\
538 & $2459740.249830 \pm 0.000490$ & -0.000345 & (d) \\
539 & $2459740.923250 \pm 0.000470$ & 0.000601 & (d) \\
540 & $2459741.595020 \pm 0.000505$ & -0.000103 & (d) \\
541 & $2459742.267420 \pm 0.000480$ & -0.000177 & (d) \\
542 & $2459742.940379 \pm 0.000126$ & 0.000308 & (c),(d)$^{\dagger}$ \\
574 & $2459764.459200 \pm 0.001700$ & -0.000042 & (b) \\
1041 & $2460078.503940 \pm 0.000200$ & -0.000706 & (c) \\
1048 & $2460083.211680 \pm 0.000120$ & -0.000285 & (c) \\
1069 & $2460097.333800 \pm 0.000130$ & -0.000121 & (c) \\
1071 & $2460098.679270 \pm 0.000130$ & 0.000401 & (c) \\
1078 & $2460103.386470 \pm 0.000130$ & 0.000282 & (c) \\
1080 & $2460104.730920 \pm 0.000150$ & -0.000216 & (c) \\
1086 & $2460108.766320 \pm 0.000130$ & 0.000339 & (c) \\
1091 & $2460112.128510 \pm 0.000130$ & 0.000159 & (c) \\
1119 & $2460130.958130 \pm 0.000750$ & 0.000504 & (b) \\
1144 & $2460147.770700 \pm 0.001300$ & 0.001222 & (b) \\
1160 & $2460158.528510 \pm 0.000545$ & -0.000554 & (b) \\
1508 & $2460392.549050 \pm 0.000970$ & -0.001000 & (b) \\
1516 & $2460397.930137 \pm 0.000486$ & 0.000294 & (e) \\
1598 & $2460453.073100 \pm 0.000460$ & 0.000381 & (d) \\
1599 & $2460453.745530 \pm 0.000460$ & 0.000337 & (d) \\
1600 & $2460454.417140 \pm 0.000500$ & -0.000527 & (d) \\
1601 & $2460455.089930 \pm 0.000460$ & -0.000211 & (d) \\
1602 & $2460455.762210 \pm 0.000475$ & -0.000405 & (d) \\
1613 & $2460463.160070 \pm 0.000510$ & 0.000240 & (d) \\
1614 & $2460463.831740 \pm 0.000460$ & -0.000564 & (d) \\
1615 & $2460464.505180 \pm 0.000485$ & 0.000402 & (d) \\
1616 & $2460465.177210 \pm 0.000400$ & -0.000042 & (d) \\
1618 & $2460466.522500 \pm 0.000495$ & 0.000299 & (d) \\
1619 & $2460467.195320 \pm 0.000460$ & 0.000645 & (d) \\
1620 & $2460467.866680 \pm 0.000490$ & -0.000469 & (d) \\
1621 & $2460468.540530 \pm 0.000470$ & 0.000907 & (d) \\
1622 & $2460469.211900 \pm 0.000490$ & -0.000197 & (d) \\
1623 & $2460469.884610 \pm 0.000505$ & 0.000039 & (d) \\
1624 & $2460470.557560 \pm 0.000465$ & 0.000515 & (d) \\
1625 & $2460471.229310 \pm 0.000505$ & -0.000209 & (d) \\
1626 & $2460471.902110 \pm 0.000480$ & 0.000117 & (d) \\
1627 & $2460472.573790 \pm 0.000455$ & -0.000678 & (d) \\
1629 & $2460473.919380 \pm 0.000570$ & -0.000036 & (d) \\
1634 & $2460477.282080 \pm 0.000550$ & 0.000294 & (d) \\
1635 & $2460477.954190 \pm 0.000515$ & -0.000070 & (d) \\
1636 & $2460478.627180 \pm 0.000470$ & 0.000446 & (d) \\
1637 & $2460479.298730 \pm 0.000505$ & -0.000479 & (d) \\
2049 & $2460756.357050 \pm 0.000260$ & -0.001487 & (f) \\
2092 & $2460785.274200 \pm 0.001150$ & -0.000723 & (f) \\
\enddata
\tablecomments{(a) WASP; (b) TFOP; (c) CHEOPS; (d) TESS; (e) HST; (f) This study. $^{\dagger}$A total of 91 light curves were collected from various sources. For epochs 523, 525, 527, and 542, TESS and CHEOPS observations occur at the same epoch; for these, we adopt the weighted average of the corresponding TESS and CHEOPS mid-transit times as a single combined value. The resulting 87 mid-transit times are listed here and used throughout the analysis.}
\end{deluxetable*}

% Normalized light curves
\begin{table*}[ht]
\centering
\caption{Normalized light curves of TOI-2109b from \textit{TESS}, \textit{CHEOPS},\textit{HST}, \textit{TFOP}, and our ground-based observations used in this analysis. Only a small portion of our ground-based data is shown here; the complete table is available in machine-readable form.}
\label{tab:flux_data}
\begin{tabular}{cccc}
\hline
\hline
Epoch&BJD$_{\mathrm{TDB}}$ & Normalized flux & Uncertainty \\
\hline
\hline
2049& 2460756.298715 & 1.000733 & 0.001223 \\
&2460756.298866 & 0.997710 & 0.001221 \\
&2460756.299016 & 1.001392 & 0.001224 \\
&-&-&-\\
2092&2460785.222097 & 1.012667 & 0.005243 \\
&2460785.222213 & 0.996880 & 0.005088 \\
&2460785.222340 & 1.008901 & 0.005191 \\
&-&-&-\\
\hline
\end{tabular}
\end{table*}

\bibliography{toi2109}{}

@ARTICLE{2021AJ....162..256W,
       author = {{Wong}, Ian and {Shporer}, Avi and {Zhou}, George and {Kitzmann}, Daniel and {Komacek}, Thaddeus D. and {Tan}, Xianyu and {Tronsgaard}, Ren{\'e} and {Buchhave}, Lars A. and {Vissapragada}, Shreyas and {Greklek-McKeon}, Michael and {Rodriguez}, Joseph E. and {Ahlers}, John P. and {Quinn}, Samuel N. and {Furlan}, Elise and {Howell}, Steve B. and {Bieryla}, Allyson and {Heng}, Kevin and {Knutson}, Heather A. and {Collins}, Karen A. and {McLeod}, Kim K. and {Berlind}, Perry and {Brown}, Peyton and {Calkins}, Michael L. and {de Leon}, Jerome P. and {Esparza-Borges}, Emma and {Esquerdo}, Gilbert A. and {Fukui}, Akihiko and {Gan}, Tianjun and {Girardin}, Eric and {Gnilka}, Crystal L. and {Ikoma}, Masahiro and {Jensen}, Eric L.~N. and {Kielkopf}, John and {Kodama}, Takanori and {Kurita}, Seiya and {Lester}, Kathryn V. and {Lewin}, Pablo and {Marino}, Giuseppe and {Murgas}, Felipe and {Narita}, Norio and {Pall{\'e}}, Enric and {Schwarz}, Richard P. and {Stassun}, Keivan G. and {Tamura}, Motohide and {Watanabe}, Noriharu and {Benneke}, Bj{\"o}rn and {Ricker}, George R. and {Latham}, David W. and {Vanderspek}, Roland and {Seager}, Sara and {Winn}, Joshua N. and {Jenkins}, Jon M. and {Caldwell}, Douglas A. and {Fong}, William and {Huang}, Chelsea X. and {Mireles}, Ismael and {Schlieder}, Joshua E. and {Shiao}, Bernie and {Noel Villase{\~n}or}, Jesus},
        title = "{TOI-2109: An Ultrahot Gas Giant on a 16 hr Orbit}",
      journal = {\aj},
         year = 2021,
        month = dec,
       volume = {162},
       number = {6},
          eid = {256},
        pages = {256},
          doi = {10.3847/1538-3881/ac26bd},
archivePrefix = {arXiv},
       eprint = {2111.12074},
 primaryClass = {astro-ph.EP},
       adsurl = {https://ui.adsabs.harvard.edu/abs/2021AJ....162..256W}
}

@MISC{2018ascl.soft12013L,
   author = {{Lightkurve Collaboration} and {Cardoso}, J.~V.~d.~M. and
             {Hedges}, C. and {Gully-Santiago}, M. and {Saunders}, N. and
             {Cody}, A.~M. and {Barclay}, T. and {Hall}, O. and
             {Sagear}, S. and {Turtelboom}, E. and {Zhang}, J. and
             {Tzanidakis}, A. and {Mighell}, K. and {Coughlin}, J. and
             {Bell}, K. and {Berta-Thompson}, Z. and {Williams}, P. and
             {Dotson}, J. and {Barentsen}, G.},
    title = "{Lightkurve: Kepler and TESS time series analysis in Python}",
howpublished = {Astrophysics Source Code Library},
     year = 2018,
    month = dec,
archivePrefix = "ascl",
   eprint = {1812.013},
   adsurl = {http://adsabs.harvard.edu/abs/2018ascl.soft12013L},
}

@ARTICLE{2020A&A...635A..24H,
       author = {{Hoyer}, S. and {Guterman}, P. and {Demangeon}, O. and {Sousa}, S.~G. and {Deleuil}, M. and {Meunier}, J.~C. and {Benz}, W.},
        title = "{Expected performances of the Characterising Exoplanet Satellite (CHEOPS). III. Data reduction pipeline: architecture and simulated performances}",
      journal = {\aap},
         year = 2020,
        month = mar,
       volume = {635},
          eid = {A24},
        pages = {A24},
          doi = {10.1051/0004-6361/201936325},
archivePrefix = {arXiv},
       eprint = {1909.08363},
 primaryClass = {astro-ph.IM},
       adsurl = {https://ui.adsabs.harvard.edu/abs/2020A&A...635A..24H}
}

@ARTICLE{2016ApJ...832..202T,
       author = {{Tsiaras}, A. and {Waldmann}, I.~P. and {Rocchetto}, M. and {Varley}, R. and {Morello}, G. and {Damiano}, M. and {Tinetti}, G.},
        title = "{A New Approach to Analyzing HST Spatial Scans: The Transmission Spectrum of HD 209458 b}",
      journal = {\apj},
         year = 2016,
        month = dec,
       volume = {832},
       number = {2},
          eid = {202},
        pages = {202},
          doi = {10.3847/0004-637X/832/2/202},
archivePrefix = {arXiv},
       eprint = {1511.07796},
 primaryClass = {astro-ph.EP},
       adsurl = {https://ui.adsabs.harvard.edu/abs/2016ApJ...832..202T}
}

@ARTICLE{2017AJ....153...77C,
       author = {{Collins}, Karen A. and {Kielkopf}, John F. and {Stassun}, Keivan G. and {Hessman}, Frederic V.},
        title = "{AstroImageJ: Image Processing and Photometric Extraction for Ultra-precise Astronomical Light Curves}",
      journal = {\aj},
         year = 2017,
        month = feb,
       volume = {153},
       number = {2},
          eid = {77},
        pages = {77},
          doi = {10.3847/1538-3881/153/2/77},
archivePrefix = {arXiv},
       eprint = {1701.04817},
 primaryClass = {astro-ph.IM},
       adsurl = {https://ui.adsabs.harvard.edu/abs/2017AJ....153...77C}
}

@ARTICLE{2013PASP..125..306F,
       author = {{Foreman-Mackey}, Daniel and {Hogg}, David W. and {Lang}, Dustin and {Goodman}, Jonathan},
        title = "{emcee: The MCMC Hammer}",
      journal = {\pasp},
         year = 2013,
        month = mar,
       volume = {125},
       number = {925},
        pages = {306},
          doi = {10.1086/670067},
archivePrefix = {arXiv},
       eprint = {1202.3665},
 primaryClass = {astro-ph.IM},
       adsurl = {https://ui.adsabs.harvard.edu/abs/2013PASP..125..306F}
}

@ARTICLE{2022ExA....53..547K,
       author = {{Kokori}, Anastasia and {Tsiaras}, Angelos and {Edwards}, Billy and {Rocchetto}, Marco and {Tinetti}, Giovanna and {W{\"u}nsche}, Ana{\"e}l and {Paschalis}, Nikolaos and {Agnihotri}, Vikrant Kumar and {Bachschmidt}, Matthieu and {Bretton}, Marc and {Caines}, Hamish and {Cal{\'o}}, Mauro and {Casali}, Roland and {Crow}, Martin and {Dawes}, Simon and {Deldem}, Marc and {Deligeorgopoulos}, Dimitrios and {Dymock}, Roger and {Evans}, Phil and {Falco}, Carmelo and {Ferratfiat}, Stephane and {Fowler}, Martin and {Futcher}, Stephen and {Guerra}, Pere and {Hurter}, Francois and {Jones}, Adrian and {Kang}, Wonseok and {Kim}, Taewoo and {Lee}, Richard and {Lopresti}, Claudio and {Marino}, Antonio and {Mallonn}, Matthias and {Mortari}, Fabio and {Morvan}, Mario and {Mugnai}, Lorenzo V. and {Nastasi}, Alessandro and {Perroud}, Val{\`e}re and {Pereira}, C{\'e}dric and {Phillips}, Mark and {Pintr}, Pavel and {Raetz}, Manfred and {Regembal}, Francois and {Savage}, John and {Sedita}, Danilo and {Sioulas}, Nick and {Strikis}, Iakovos and {Thurston}, Geoffrey and {Tomacelli}, Andrea and {Tomatis}, Alberto},
        title = "{ExoClock project: an open platform for monitoring the ephemerides of Ariel targets with contributions from the public}",
      journal = {Experimental Astronomy},
         year = 2022,
        month = apr,
       volume = {53},
       number = {2},
        pages = {547-588},
          doi = {10.1007/s10686-020-09696-3},
archivePrefix = {arXiv},
       eprint = {2012.07478},
 primaryClass = {astro-ph.EP},
       adsurl = {https://ui.adsabs.harvard.edu/abs/2022ExA....53..547K}
}

@ARTICLE{2018ARA&A..56..175D,
       author = {{Dawson}, Rebekah I. and {Johnson}, John Asher},
        title = "{Origins of Hot Jupiters}",
      journal = {\araa},
         year = 2018,
        month = sep,
       volume = {56},
        pages = {175-221},
          doi = {10.1146/annurev-astro-081817-051853},
archivePrefix = {arXiv},
       eprint = {1801.06117},
 primaryClass = {astro-ph.EP},
       adsurl = {https://ui.adsabs.harvard.edu/abs/2018ARA&A..56..175D}
}

@ARTICLE{2025ApJ...988...66A,
       author = {{Alvarado-Montes}, Jaime A. and {Sucerquia}, Mario and {Zuluaga}, Jorge I. and {Schwab}, Christian},
        title = "{Orbital Decay of the Ultra-hot Jupiter TOI-2109b: Tidal Constraints and Transit-timing Analysis}",
      journal = {\apj},
         year = 2025,
        month = jul,
       volume = {988},
       number = {1},
          eid = {66},
        pages = {66},
          doi = {10.3847/1538-4357/ade057},
archivePrefix = {arXiv},
       eprint = {2505.18941},
 primaryClass = {astro-ph.EP},
       adsurl = {https://ui.adsabs.harvard.edu/abs/2025ApJ...988...66A}
}

@ARTICLE{2012A&A...537A.128R,
       author = {{Rein}, H. and {Liu}, S.-F.},
        title = "{REBOUND: an open-source multi-purpose N-body code for collisional dynamics}",
      journal = {\aap},
         year = 2012,
        month = jan,
       volume = {537},
          eid = {A128},
        pages = {A128},
          doi = {10.1051/0004-6361/201118085},
archivePrefix = {arXiv},
       eprint = {1110.4876},
 primaryClass = {astro-ph.EP},
       adsurl = {https://ui.adsabs.harvard.edu/abs/2012A&A...537A.128R}
}

@ARTICLE{2000A&AS..147..205C,
       author = {{Cincotta}, P.~M. and {Sim{\'o}}, C.},
        title = "{Simple tools to study global dynamics in non-axisymmetric galactic potentials - I}",
      journal = {\aaps},
         year = 2000,
        month = dec,
       volume = {147},
        pages = {205-228},
          doi = {10.1051/aas:2000108},
       adsurl = {https://ui.adsabs.harvard.edu/abs/2000A&AS..147..205C}
}

@ARTICLE{2015MNRAS.452..376R,
       author = {{Rein}, Hanno and {Tamayo}, Daniel},
        title = "{WHFAST: a fast and unbiased implementation of a symplectic Wisdom-Holman integrator for long-term gravitational simulations}",
      journal = {\mnras},
         year = 2015,
        month = sep,
       volume = {452},
       number = {1},
        pages = {376-388},
          doi = {10.1093/mnras/stv1257},
archivePrefix = {arXiv},
       eprint = {1506.01084},
 primaryClass = {astro-ph.EP},
       adsurl = {https://ui.adsabs.harvard.edu/abs/2015MNRAS.452..376R}
}

@ARTICLE{2011IJNLM..46...23M,
       author = {{Maffione}, N.~P. and {Giordano}, C.~M. and {Cincotta}, P.~M.},
        title = "{Testing a fast dynamical indicator: The MEGNO}",
      journal = {International Journal of Non Linear Mechanics},
         year = 2011,
        month = jan,
       volume = {46},
       number = {1},
        pages = {23-34},
          doi = {10.1016/j.ijnonlinmec.2010.06.008},
archivePrefix = {arXiv},
       eprint = {1108.5481},
 primaryClass = {nlin.CD},
       adsurl = {https://ui.adsabs.harvard.edu/abs/2011IJNLM..46...23M}
}

@ARTICLE{2012ApJ...761..122L,
       author = {{Lithwick}, Yoram and {Xie}, Jiwei and {Wu}, Yanqin},
        title = "{Extracting Planet Mass and Eccentricity from TTV Data}",
      journal = {\apj},
         year = 2012,
        month = dec,
       volume = {761},
       number = {2},
          eid = {122},
        pages = {122},
          doi = {10.1088/0004-637X/761/2/122},
archivePrefix = {arXiv},
       eprint = {1207.4192},
 primaryClass = {astro-ph.EP},
       adsurl = {https://ui.adsabs.harvard.edu/abs/2012ApJ...761..122L}
}

@BOOK{1999ssd..book.....M,
       author = {{Murray}, Carl D. and {Dermott}, Stanley F.},
        title = "{Solar System Dynamics}",
         year = 1999,
          doi = {10.1017/CBO9781139174817},
       adsurl = {https://ui.adsabs.harvard.edu/abs/1999ssd..book.....M}
}

@ARTICLE{2004astro.ph.12032A,
       author = {{Agol}, E. and {Steffen}, J. and {Sari}, R. and {Clarkson}, W.},
        title = "{On detecting terrestrial planets with timing of giant planet transits}",
      journal = {arXiv e-prints},
         year = 2004,
        month = dec,
          eid = {astro-ph/0412032},
        pages = {astro-ph/0412032},
          doi = {10.48550/arXiv.astro-ph/0412032},
archivePrefix = {arXiv},
       eprint = {astro-ph/0412032},
 primaryClass = {astro-ph},
       adsurl = {https://ui.adsabs.harvard.edu/abs/2004astro.ph.12032A}
}

@ARTICLE{1997JGOpt..11..341S,
       author = {{Storn}, Rainer and {Price}, Kenneth},
        title = "{Differential Evolution - A Simple and Efficient Heuristic for global Optimization over Continuous Spaces}",
      journal = {Journal of Global Optimization},
         year = 1997,
        month = dec,
       volume = {11},
        pages = {341-359},
          doi = {10.1023/A:1008202821328},
       adsurl = {https://ui.adsabs.harvard.edu/abs/1997JGOpt..11..341S}
}

@ARTICLE{Harre2024AA,
       author = {{Harre}, J.-V. and {Smith}, A.~M.~S. and {Barros}, S.~C.~C. and {Singh}, V. and {Korth}, J. and {Brandeker}, A. and {Collier Cameron}, A. and {Lendl}, M. and {Wilson}, T.~G. and {Borsato}, L. and {Csizmadia}, Sz. and {Cabrera}, J. and {Parviainen}, H. and {Correia}, A.~C.~M. and {Akinsanmi}, B. and {Rosario}, N. and {Leonardi}, P. and {Serrano}, L.~M. and {Alibert}, Y. and {Alonso}, R. and {Asquier}, J. and {B{\'a}rczy}, T. and {Barrado Navascues}, D. and {Baumjohann}, W. and {Benz}, W. and {Billot}, N. and {Broeg}, C. and {Busch}, M.-D. and {Cubillos}, P.~E. and {Davies}, M.~B. and {Deleuil}, M. and {Deline}, A. and {Delrez}, L. and {Demangeon}, O.~D.~S. and {Demory}, B.-O. and {Derekas}, A. and {Edwards}, B. and {Ehrenreich}, D. and {Erikson}, A. and {Fortier}, A. and {Fossati}, L. and {Fridlund}, M. and {Gandolfi}, D. and {Gazeas}, K. and {Gillon}, M. and {G{\"u}del}, M. and {G{\"u}nther}, M.~N. and {Heitzmann}, A. and {Helling}, Ch. and {Isaak}, K.~G. and {Kiss}, L.~L. and {Lam}, K.~W.~F. and {Laskar}, J. and {Lecavelier des Etangs}, A. and {Magrin}, D. and {Maxted}, P.~F.~L. and {Mer{\'\i}n}, B. and {Mordasini}, C. and {Nascimbeni}, V. and {Olofsson}, G. and {Ottensamer}, R. and {Pagano}, I. and {Pall{\'e}}, E. and {Peter}, G. and {Piazza}, D. and {Piotto}, G. and {Pollacco}, D. and {Queloz}, D. and {Ragazzoni}, R. and {Rando}, N. and {Rauer}, H. and {Ribas}, I. and {Santos}, N.~C. and {Scandariato}, G. and {S{\'e}gransan}, D. and {Simon}, A.~E. and {Sousa}, S.~G. and {Stalport}, M. and {Sulis}, S. and {Szab{\'o}}, Gy. M. and {Udry}, S. and {Ulmer}, B. and {Van Grootel}, V. and {Venturini}, J. and {Villaver}, E. and {Viotto}, V. and {Walton}, N.~A. and {West}, R. and {Westerdorff}, K.},
        title = "{Hints of a close outer companion to the ultra-hot Jupiter TOI-2109 b}",
      journal = {\aap},
         year = 2024,
        month = dec,
       volume = {692},
          eid = {A254},
        pages = {A254},
          doi = {10.1051/0004-6361/202451068},
archivePrefix = {arXiv},
       eprint = {2411.07797},
 primaryClass = {astro-ph.EP},
       adsurl = {https://ui.adsabs.harvard.edu/abs/2024A&A...692A.254H}
}

@ARTICLE{1995Natur.378..355M,
       author = {{Mayor}, Michel and {Queloz}, Didier},
        title = "{A Jupiter-mass companion to a solar-type star}",
      journal = {\nat},
         year = 1995,
        month = nov,
       volume = {378},
       number = {6555},
        pages = {355-359},
          doi = {10.1038/378355a0},
       adsurl = {https://ui.adsabs.harvard.edu/abs/1995Natur.378..355M}
}

@ARTICLE{2015ARA&A..53..409W,
       author = {{Winn}, Joshua N. and {Fabrycky}, Daniel C.},
        title = "{The Occurrence and Architecture of Exoplanetary Systems}",
      journal = {\araa},
         year = 2015,
        month = aug,
       volume = {53},
        pages = {409-447},
          doi = {10.1146/annurev-astro-082214-122246},
archivePrefix = {arXiv},
       eprint = {1410.4199},
 primaryClass = {astro-ph.EP},
       adsurl = {https://ui.adsabs.harvard.edu/abs/2015ARA&A..53..409W}
}

@ARTICLE{2012A&A...547A.111M,
       author = {{Mordasini}, C. and {Alibert}, Y. and {Klahr}, H. and {Henning}, T.},
        title = "{Characterization of exoplanets from their formation. I. Models of combined planet formation and evolution}",
      journal = {\aap},
         year = 2012,
        month = nov,
       volume = {547},
          eid = {A111},
        pages = {A111},
          doi = {10.1051/0004-6361/201118457},
archivePrefix = {arXiv},
       eprint = {1206.6103},
 primaryClass = {astro-ph.EP},
       adsurl = {https://ui.adsabs.harvard.edu/abs/2012A&A...547A.111M}
}

@ARTICLE{2019AJ....157..180P,
       author = {{Petrovich}, Cristobal and {Deibert}, Emily and {Wu}, Yanqin},
        title = "{Ultra-short-period Planets from Secular Chaos}",
      journal = {\aj},
         year = 2019,
        month = may,
       volume = {157},
       number = {5},
          eid = {180},
        pages = {180},
          doi = {10.3847/1538-3881/ab0e0a},
archivePrefix = {arXiv},
       eprint = {1804.05065},
 primaryClass = {astro-ph.EP},
       adsurl = {https://ui.adsabs.harvard.edu/abs/2019AJ....157..180P}
}

@ARTICLE{2007ApJ...661.1180O,
       author = {{Ogilvie}, G.~I. and {Lin}, D.~N.~C.},
        title = "{Tidal Dissipation in Rotating Solar-Type Stars}",
      journal = {\apj},
         year = 2007,
        month = jun,
       volume = {661},
       number = {2},
        pages = {1180-1191},
          doi = {10.1086/515435},
archivePrefix = {arXiv},
       eprint = {astro-ph/0702492},
 primaryClass = {astro-ph},
       adsurl = {https://ui.adsabs.harvard.edu/abs/2007ApJ...661.1180O}
}

@ARTICLE{2015ApJ...802..116D,
       author = {{Deck}, Katherine M. and {Agol}, Eric},
        title = "{Measurement of Planet Masses with Transit Timing Variations Due to Synodic {\textquotedblleft}Chopping{\textquotedblright} Effects}",
      journal = {\apj},
         year = 2015,
        month = apr,
       volume = {802},
       number = {2},
          eid = {116},
        pages = {116},
          doi = {10.1088/0004-637X/802/2/116},
archivePrefix = {arXiv},
       eprint = {1411.0004},
 primaryClass = {astro-ph.EP},
       adsurl = {https://ui.adsabs.harvard.edu/abs/2015ApJ...802..116D}
}

@ARTICLE{2014AJ....147...32G,
       author = {{Goldreich}, Peter and {Schlichting}, Hilke E.},
        title = "{Overstable Librations can Account for the Paucity of Mean Motion Resonances among Exoplanet Pairs}",
      journal = {\aj},
         year = 2014,
        month = feb,
       volume = {147},
       number = {2},
          eid = {32},
        pages = {32},
          doi = {10.1088/0004-6256/147/2/32},
archivePrefix = {arXiv},
       eprint = {1308.4688},
 primaryClass = {astro-ph.EP},
       adsurl = {https://ui.adsabs.harvard.edu/abs/2014AJ....147...32G}
}

@ARTICLE{2025AJ....169..243J,
       author = {{Jiang}, Ing-Guey and {Yeh}, Li-Chin and {Edwards}, Billy and {Yang}, Ming and {Stassun}, Keivan G. and {A-thano}, Napaporn},
        title = "{Toward High-precision Mass Measurements of Two Sub-Neptunes in the K2-266 Planetary System Through Transit Timing}",
      journal = {\aj},
         year = 2025,
        month = may,
       volume = {169},
       number = {5},
          eid = {243},
        pages = {243},
          doi = {10.3847/1538-3881/adbe7a},
archivePrefix = {arXiv},
       eprint = {2503.08905},
 primaryClass = {astro-ph.EP},
       adsurl = {https://ui.adsabs.harvard.edu/abs/2025AJ....169..243J}
}

@ARTICLE{2016AJ....151...17J,
       author = {{Jiang}, Ing-Guey and {Lai}, Chien-Yo and {Savushkin}, Alexander and {Mkrtichian}, David and {Antonyuk}, Kirill and {Griv}, Evgeny and {Hsieh}, He-Feng and {Yeh}, Li-Chin},
        title = "{The Possible Orbital Decay and Transit Timing Variations of the Planet WASP-43b}",
      journal = {\aj},
         year = 2016,
        month = jan,
       volume = {151},
       number = {1},
          eid = {17},
        pages = {17},
          doi = {10.3847/0004-6256/151/1/17},
archivePrefix = {arXiv},
       eprint = {1511.00768},
 primaryClass = {astro-ph.EP},
       adsurl = {https://ui.adsabs.harvard.edu/abs/2016AJ....151...17J}
}

@ARTICLE{2024NewA..10602130Y,
       author = {{Yeh}, Li-Chin and {Jiang}, Ing-Guey and {A-thano}, Napaporn},
        title = "{Searching for candidates of orbital decays among transit exoplanets}",
      journal = {\na},
         year = 2024,
        month = feb,
       volume = {106},
          eid = {102130},
        pages = {102130},
          doi = {10.1016/j.newast.2023.102130},
archivePrefix = {arXiv},
       eprint = {2310.08953},
 primaryClass = {astro-ph.EP},
       adsurl = {https://ui.adsabs.harvard.edu/abs/2024NewA..10602130Y}
}

@ARTICLE{2026AJ....171..321A,
       author = {{A-thano}, Napaporn and {Awiphan}, Supachai and {Kerins}, Eamonn and {Priyadarshi}, Akshay and {McDonald}, Iain and {Jiang}, Ing-Guey and {Joshi}, Yogesh C. and {Yang}, Fan and {Janiak}, Ida and {Munsaket}, Patcharawee and {Qadir}, Yasir Abdul and {Rattanamala}, Ronnakrit and {Tasuya}, Orarik and {Boonsoy}, Ekburus and {Sanguansak}, Nuanwan and {Rattanasai}, Rattiyakorn and {Padjaroen}, Thammasorn and {Komonjinda}, Siramas and {Pichadee}, Sawatkamol and {Sudsap}, Ananpol and {Chandaiam}, Smanchan and {Choonhakit}, Boonyarit and {Wutsang}, Suwanit and {Dhillon}, Vik S.},
        title = "{The Transit Timing and Transmission Spectrum of Hot Jupiter WASP-43 b from a Decade of Multiband Transit follow-up Observations}",
      journal = {\aj},
         year = 2026,
        month = may,
       volume = {171},
       number = {5},
          eid = {321},
        pages = {321},
          doi = {10.3847/1538-3881/ae5b78},
archivePrefix = {arXiv},
       eprint = {2604.05907},
 primaryClass = {astro-ph.EP},
       adsurl = {https://ui.adsabs.harvard.edu/abs/2026AJ....171..321A}
}

@ARTICLE{2011A&A...529A..50L,
       author = {{Lanza}, A.~F. and {Damiani}, C. and {Gandolfi}, D.},
        title = "{Constraining tidal dissipation in F-type main-sequence stars: the case of CoRoT-11}",
      journal = {\aap},
         year = 2011,
        month = may,
       volume = {529},
          eid = {A50},
        pages = {A50},
          doi = {10.1051/0004-6361/201016144},
archivePrefix = {arXiv},
       eprint = {1012.5791},
 primaryClass = {astro-ph.SR},
       adsurl = {https://ui.adsabs.harvard.edu/abs/2011A&A...529A..50L}
}

@ARTICLE{2006PASP..118.1407P,
       author = {{Pollacco}, D.~L. and {Skillen}, I. and {Collier Cameron}, A. and {Christian}, D.~J. and {Hellier}, C. and {Irwin}, J. and {Lister}, T.~A. and {Street}, R.~A. and {West}, R.~G. and {Anderson}, D.~R. and {Clarkson}, W.~I. and {Deeg}, H. and {Enoch}, B. and {Evans}, A. and {Fitzsimmons}, A. and {Haswell}, C.~A. and {Hodgkin}, S. and {Horne}, K. and {Kane}, S.~R. and {Keenan}, F.~P. and {Maxted}, P.~F.~L. and {Norton}, A.~J. and {Osborne}, J. and {Parley}, N.~R. and {Ryans}, R.~S.~I. and {Smalley}, B. and {Wheatley}, P.~J. and {Wilson}, D.~M.},
        title = "{The WASP Project and the SuperWASP Cameras}",
      journal = {\pasp},
         year = 2006,
        month = oct,
       volume = {118},
       number = {848},
        pages = {1407-1418},
          doi = {10.1086/508556},
archivePrefix = {arXiv},
       eprint = {astro-ph/0608454},
 primaryClass = {astro-ph},
       adsurl = {https://ui.adsabs.harvard.edu/abs/2006PASP..118.1407P}
}

@ARTICLE{2017AJ....154....4P,
       author = {{Patra}, Kishore C. and {Winn}, Joshua N. and {Holman}, Matthew J. and {Yu}, Liang and {Deming}, Drake and {Dai}, Fei},
        title = "{The Apparently Decaying Orbit of WASP-12b}",
      journal = {\aj},
         year = 2017,
        month = jul,
       volume = {154},
       number = {1},
          eid = {4},
        pages = {4},
          doi = {10.3847/1538-3881/aa6d75},
archivePrefix = {arXiv},
       eprint = {1703.06582},
 primaryClass = {astro-ph.EP},
       adsurl = {https://ui.adsabs.harvard.edu/abs/2017AJ....154....4P}
}

@ARTICLE{2025NewA..11902390P,
       author = {{Parthasarathy}, Kaviya and {Liu}, Hsin-Min and {Jiang}, Ing-Guey and {Yeh}, Li-Chin and {A-thano}, Napaporn and {Awiphan}, Supachai and {Cheng}, Wen-Chi and {Sariya}, Devesh P. and {Biswas}, Shraddha and {Bisht}, Devendra and {Griv}, Evgeny and {Mkrtichian}, David and {Mannaday}, Vineet Kumar and {Thakur}, Parijat and {Shlyapnikov}, Aleksey},
        title = "{Transit timing variations of the sub-Saturn exoplanet HAT-P-12b}",
      journal = {\na},
         year = 2025,
        month = oct,
       volume = {119},
          eid = {102390},
        pages = {102390},
          doi = {10.1016/j.newast.2025.102390},
archivePrefix = {arXiv},
       eprint = {2503.10441},
 primaryClass = {astro-ph.EP},
       adsurl = {https://ui.adsabs.harvard.edu/abs/2025NewA..11902390P}
}

@ARTICLE{2018AcA....68..371M,
       author = {{Maciejewski}, G. and {Fern{\'a}ndez}, M. and {Aceituno}, F. and {Mart{\'\i}n-Ruiz}, S. and {Ohlert}, J. and {Dimitrov}, D. and {Szyszka}, K. and {von Essen}, C. and {Mugrauer}, M. and {Bischoff}, R. and {Michel}, K.-U. and {Mallonn}, M. and {Stangret}, M. and {Mo{\'z}dzierski}, D.},
        title = "{Planet-Star Interactions with Precise Transit Timing. I. The Refined Orbital Decay Rate for WASP-12 b and Initial Constraints for HAT-P-23 b, KELT-1 b, KELT-16 b, WASP-33 b and WASP-103 b}",
      journal = {\actaa},
         year = 2018,
        month = dec,
       volume = {68},
       number = {4},
        pages = {371-401},
          doi = {10.32023/0001-5237/68.4.4},
archivePrefix = {arXiv},
       eprint = {1812.02438},
 primaryClass = {astro-ph.EP},
       adsurl = {https://ui.adsabs.harvard.edu/abs/2018AcA....68..371M}
}

@INPROCEEDINGS{2009IAUS..253..163W,
       author = {{Wolf}, Aaron S. and {Ragozzine}, Darin},
        title = "{Probing the Interiors of Very Hot Jupiters Using Transit Light Curves}",
    booktitle = {Transiting Planets},
         year = 2009,
       editor = {{Pont}, Fr{\'e}d{\'e}ric and {Sasselov}, Dimitar and {Holman}, Matthew J.},
       series = {IAU Symposium},
       volume = {253},
        month = feb,
        pages = {163-169},
          doi = {10.1017/S1743921308026367},
       adsurl = {https://ui.adsabs.harvard.edu/abs/2009IAUS..253..163W}
}

@ARTICLE{1995Ap&SS.226...99G,
       author = {{Gim{\'e}nez}, A. and {Bastero}, M.},
        title = "{A Revision of the Ephemeris-Curve Equations for Eclipsing Binaries with Apsidal Motion}",
      journal = {\apss},
         year = 1995,
        month = apr,
       volume = {226},
       number = {1},
        pages = {99-107},
          doi = {10.1007/BF00626903},
       adsurl = {https://ui.adsabs.harvard.edu/abs/1995Ap&SS.226...99G}
}

@ARTICLE{2019A&A...628A.116V,
       author = {{von Essen}, C. and {Wedemeyer}, S. and {Sosa}, M.~S. and {Hjorth}, M. and {Parkash}, V. and {Freudenthal}, J. and {Mallonn}, M. and {Micul{\'a}n}, R.~G. and {Zibecchi}, L. and {Cellone}, S. and {Torres}, A.~F.},
        title = "{Indications for transit-timing variations in the exo-Neptune HAT-P-26b}",
      journal = {\aap},
         year = 2019,
        month = aug,
       volume = {628},
          eid = {A116},
        pages = {A116},
          doi = {10.1051/0004-6361/201731966},
archivePrefix = {arXiv},
       eprint = {1904.06360},
 primaryClass = {astro-ph.EP},
       adsurl = {https://ui.adsabs.harvard.edu/abs/2019A&A...628A.116V}
}

@ARTICLE{1976Ap&SS..39..447L,
       author = {{Lomb}, N.~R.},
        title = "{Least-Squares Frequency Analysis of Unequally Spaced Data}",
      journal = {\apss},
         year = 1976,
        month = feb,
       volume = {39},
       number = {2},
        pages = {447-462},
          doi = {10.1007/BF00648343},
       adsurl = {https://ui.adsabs.harvard.edu/abs/1976Ap&SS..39..447L}
}

@ARTICLE{1982ApJ...263..835S,
       author = {{Scargle}, J.~D.},
        title = "{Studies in astronomical time series analysis. II. Statistical aspects of spectral analysis of unevenly spaced data.}",
      journal = {\apj},
         year = 1982,
        month = dec,
       volume = {263},
        pages = {835-853},
          doi = {10.1086/160554},
       adsurl = {https://ui.adsabs.harvard.edu/abs/1982ApJ...263..835S}
}

@ARTICLE{2014ApJ...787..132D,
       author = {{Deck}, Katherine M. and {Agol}, Eric and {Holman}, Matthew J. and {Nesvorn{\'y}}, David},
        title = "{TTVFast: An Efficient and Accurate Code for Transit Timing Inversion Problems}",
      journal = {\apj},
         year = 2014,
        month = jun,
       volume = {787},
       number = {2},
          eid = {132},
        pages = {132},
          doi = {10.1088/0004-637X/787/2/132},
archivePrefix = {arXiv},
       eprint = {1403.1895},
 primaryClass = {astro-ph.EP},
       adsurl = {https://ui.adsabs.harvard.edu/abs/2014ApJ...787..132D}
}

@MISC{2019ascl.soft06010C,
  author = {{Czesla}, Stefan and {Schr{\"o}ter}, Sebastian and {Schneider}, Christian P. and {Huber}, Klaus F. and {Pfeifer}, Fabian and {Andreasen}, Daniel T. and {Zechmeister}, Mathias},
  title = "{PyA: Python astronomy-related packages}",
  howpublished = {Astrophysics Source Code Library, record ascl:1906.010},
  year = 2019,
  month = jun
}

@article{virtanen2020scipy,
  title={SciPy 1.0: fundamental algorithms for scientific computing in Python},
  author={Virtanen, Pauli and Gommers, Ralf and Oliphant, Travis E and Haberland, Matt and Reddy, Tyler and Cournapeau, David and Burovski, Evgeni and Peterson, Pearu and Weckesser, Eric and Bright, Jonathan and others},
  journal={Nature Methods},
  volume={17},
  number={3},
  pages={261--272},
  year={2020},
  publisher={Nature Publishing Group}
}

@ARTICLE{2010arXiv1008.4686H,
       author = {{Hogg}, David W. and {Bovy}, Jo and {Lang}, Dustin},
        title = "{Data analysis recipes: Fitting a model to data}",
      journal = {arXiv e-prints},
         year = 2010,
        month = aug,
          eid = {arXiv:1008.4686},
        pages = {arXiv:1008.4686},
          doi = {10.48550/arXiv.1008.4686},
archivePrefix = {arXiv},
       eprint = {1008.4686},
 primaryClass = {astro-ph.IM},
       adsurl = {https://ui.adsabs.harvard.edu/abs/2010arXiv1008.4686H}
}

@ARTICLE{Brown:2013,
       author = {{Brown}, T.~M. and {Baliber}, N. and {Bianco}, F.~B. and {Bowman}, M. and {Burleson}, B. and {Conway}, P. and {Crellin}, M. and {Depagne}, {\'E}. and {De Vera}, J. and {Dilday}, B. and {Dragomir}, D. and {Dubberley}, M. and {Eastman}, J.~D. and {Elphick}, M. and {Falarski}, M. and {Foale}, S. and {Ford}, M. and {Fulton}, B.~J. and {Garza}, J. and {Gomez}, E.~L. and {Graham}, M. and {Greene}, R. and {Haldeman}, B. and {Hawkins}, E. and {Haworth}, B. and {Haynes}, R. and {Hidas}, M. and {Hjelstrom}, A.~E. and {Howell}, D.~A. and {Hygelund}, J. and {Lister}, T.~A. and {Lobdill}, R. and {Martinez}, J. and {Mullins}, D.~S. and {Norbury}, M. and {Parrent}, J. and {Paulson}, R. and {Petry}, D.~L. and {Pickles}, A. and {Posner}, V. and {Rosing}, W.~E. and {Ross}, R. and {Sand}, D.~J. and {Saunders}, E.~S. and {Shobbrook}, J. and {Shporer}, A. and {Street}, R.~A. and {Thomas}, D. and {Tsapras}, Y. and {Tufts}, J.~R. and {Valenti}, S. and {Vander Horst}, K. and {Walker}, Z. and {White}, G. and {Willis}, M.},
        title = "{Las Cumbres Observatory Global Telescope Network}",
      journal = {\pasp},
         year = 2013,
        month = sep,
       volume = {125},
       number = {931},
        pages = {1031},
          doi = {10.1086/673168},
archivePrefix = {arXiv},
       eprint = {1305.2437},
 primaryClass = {astro-ph.IM},
       adsurl = {https://ui.adsabs.harvard.edu/abs/2013PASP..125.1031B}
}

@INPROCEEDINGS{McCully:2018,
       author = {{McCully}, Curtis and {Volgenau}, Nikolaus H. and
         {Harbeck}, Daniel-Rolf and {Lister}, Tim A. and {Saunders}, Eric S. and
         {Turner}, Monica L. and {Siiverd}, Robert J. and {Bowman}, Mark},
        title = "{Real-time processing of the imaging data from the network of Las Cumbres Observatory Telescopes using BANZAI}",
    booktitle = {\procspie},
         year = "2018",
       series = {Society of Photo-Optical Instrumentation Engineers (SPIE) Conference Series},
       volume = {10707},
        month = "Jul",
          eid = {107070K},
        pages = {107070K},
          doi = {10.1117/12.2314340},
archivePrefix = {arXiv},
       eprint = {1811.04163},
 primaryClass = {astro-ph.IM},
       adsurl = {https://ui.adsabs.harvard.edu/abs/2018SPIE10707E..0KM}
}

@ARTICLE{2018ApJ...857L..20B,
       author = {{Bell}, Taylor J. and {Cowan}, Nicolas B.},
        title = "{Increased Heat Transport in Ultra-hot Jupiter Atmospheres through H$_{2}$ Dissociation and Recombination}",
      journal = {\apjl},
         year = 2018,
        month = apr,
       volume = {857},
       number = {2},
          eid = {L20},
        pages = {L20},
          doi = {10.3847/2041-8213/aabcc8},
archivePrefix = {arXiv},
       eprint = {1802.07725},
 primaryClass = {astro-ph.EP},
       adsurl = {https://ui.adsabs.harvard.edu/abs/2018ApJ...857L..20B}
}

@ARTICLE{1973ApJ...180..307C,
       author = {{Counselman}, III, Charles C.},
        title = "{Outcomes of Tidal Evolution}",
      journal = {\apj},
         year = 1973,
        month = feb,
       volume = {180},
        pages = {307-316},
          doi = {10.1086/151964},
       adsurl = {https://ui.adsabs.harvard.edu/abs/1973ApJ...180..307C}
}

@ARTICLE{1996Sci...274..954R,
       author = {{Rasio}, Frederic A. and {Ford}, Eric B.},
        title = "{Dynamical instabilities and the formation of extrasolar planetary systems}",
      journal = {Science},
         year = 1996,
        month = nov,
       volume = {274},
        pages = {954-956},
          doi = {10.1126/science.274.5289.954},
       adsurl = {https://ui.adsabs.harvard.edu/abs/1996Sci...274..954R}
}

@ARTICLE{2020ApJ...888L...5Y,
       author = {{Yee}, Samuel W. and {Winn}, Joshua N. and {Knutson}, Heather A. and {Patra}, Kishore C. and {Vissapragada}, Shreyas and {Zhang}, Michael M. and {Holman}, Matthew J. and {Shporer}, Avi and {Wright}, Jason T.},
        title = "{The Orbit of WASP-12b Is Decaying}",
      journal = {\apjl},
         year = 2020,
        month = jan,
       volume = {888},
       number = {1},
          eid = {L5},
        pages = {L5},
          doi = {10.3847/2041-8213/ab5c16},
archivePrefix = {arXiv},
       eprint = {1911.09131},
 primaryClass = {astro-ph.EP},
       adsurl = {https://ui.adsabs.harvard.edu/abs/2020ApJ...888L...5Y}
}

@ARTICLE{2020MNRAS.498.2270B,
       author = {{Barker}, A.~J.},
        title = "{Tidal dissipation in evolving low-mass and solar-type stars with predictions for planetary orbital decay}",
      journal = {\mnras},
         year = 2020,
        month = oct,
       volume = {498},
       number = {2},
        pages = {2270-2294},
          doi = {10.1093/mnras/staa2405},
archivePrefix = {arXiv},
       eprint = {2008.03262},
 primaryClass = {astro-ph.EP},
       adsurl = {https://ui.adsabs.harvard.edu/abs/2020MNRAS.498.2270B}
}

@ARTICLE{2020AJ....159...75M,
       author = {{Morello}, G. and {Claret}, A. and {Martin-Lagarde}, M. and {Cossou}, C. and {Tsiaras}, A. and {Lagage}, P.-O.},
        title = "{The ExoTETHyS Package: Tools for Exoplanetary Transits around Host Stars}",
      journal = {\aj},
         year = 2020,
        month = feb,
       volume = {159},
       number = {2},
          eid = {75},
        pages = {75},
          doi = {10.3847/1538-3881/ab63dc},
archivePrefix = {arXiv},
       eprint = {1908.09599},
 primaryClass = {astro-ph.EP},
       adsurl = {https://ui.adsabs.harvard.edu/abs/2020AJ....159...75M}
}

@ARTICLE{2013ApJ...766...81F,
       author = {{Fressin}, Fran{\c{c}}ois and {Torres}, Guillermo and {Charbonneau}, David and {Bryson}, Stephen T. and {Christiansen}, Jessie and {Dressing}, Courtney D. and {Jenkins}, Jon M. and {Walkowicz}, Lucianne M. and {Batalha}, Natalie M.},
        title = "{The False Positive Rate of Kepler and the Occurrence of Planets}",
      journal = {\apj},
         year = 2013,
        month = apr,
       volume = {766},
       number = {2},
          eid = {81},
        pages = {81},
          doi = {10.1088/0004-637X/766/2/81},
archivePrefix = {arXiv},
       eprint = {1301.0842},
 primaryClass = {astro-ph.EP},
       adsurl = {https://ui.adsabs.harvard.edu/abs/2013ApJ...766...81F}
}

@ARTICLE{2018AJ....155...89P,
       author = {{Petigura}, Erik A. and {Marcy}, Geoffrey W. and {Winn}, Joshua N. and {Weiss}, Lauren M. and {Fulton}, Benjamin J. and {Howard}, Andrew W. and {Sinukoff}, Evan and {Isaacson}, Howard and {Morton}, Timothy D. and {Johnson}, John Asher},
        title = "{The California-Kepler Survey. IV. Metal-rich Stars Host a Greater Diversity of Planets}",
      journal = {\aj},
         year = 2018,
        month = feb,
       volume = {155},
       number = {2},
          eid = {89},
        pages = {89},
          doi = {10.3847/1538-3881/aaa54c},
archivePrefix = {arXiv},
       eprint = {1712.04042},
 primaryClass = {astro-ph.EP},
       adsurl = {https://ui.adsabs.harvard.edu/abs/2018AJ....155...89P}
}

@ARTICLE{2012ApJS..201...15H,
       author = {{Howard}, Andrew W. and {Marcy}, Geoffrey W. and {Bryson}, Stephen T. and {Jenkins}, Jon M. and {Rowe}, Jason F. and {Batalha}, Natalie M. and {Borucki}, William J. and {Koch}, David G. and {Dunham}, Edward W. and {Gautier}, III, Thomas N. and {Van Cleve}, Jeffrey and {Cochran}, William D. and {Latham}, David W. and {Lissauer}, Jack J. and {Torres}, Guillermo and {Brown}, Timothy M. and {Gilliland}, Ronald L. and {Buchhave}, Lars A. and {Caldwell}, Douglas A. and {Christensen-Dalsgaard}, J{\o}rgen and {Ciardi}, David and {Fressin}, Francois and {Haas}, Michael R. and {Howell}, Steve B. and {Kjeldsen}, Hans and {Seager}, Sara and {Rogers}, Leslie and {Sasselov}, Dimitar D. and {Steffen}, Jason H. and {Basri}, Gibor S. and {Charbonneau}, David and {Christiansen}, Jessie and {Clarke}, Bruce and {Dupree}, Andrea and {Fabrycky}, Daniel C. and {Fischer}, Debra A. and {Ford}, Eric B. and {Fortney}, Jonathan J. and {Tarter}, Jill and {Girouard}, Forrest R. and {Holman}, Matthew J. and {Johnson}, John Asher and {Klaus}, Todd C. and {Machalek}, Pavel and {Moorhead}, Althea V. and {Morehead}, Robert C. and {Ragozzine}, Darin and {Tenenbaum}, Peter and {Twicken}, Joseph D. and {Quinn}, Samuel N. and {Isaacson}, Howard and {Shporer}, Avi and {Lucas}, Philip W. and {Walkowicz}, Lucianne M. and {Welsh}, William F. and {Boss}, Alan and {Devore}, Edna and {Gould}, Alan and {Smith}, Jeffrey C. and {Morris}, Robert L. and {Prsa}, Andrej and {Morton}, Timothy D. and {Still}, Martin and {Thompson}, Susan E. and {Mullally}, Fergal and {Endl}, Michael and {MacQueen}, Phillip J.},
        title = "{Planet Occurrence within 0.25 AU of Solar-type Stars from Kepler}",
      journal = {\apjs},
         year = 2012,
        month = aug,
       volume = {201},
       number = {2},
          eid = {15},
        pages = {15},
          doi = {10.1088/0067-0049/201/2/15},
archivePrefix = {arXiv},
       eprint = {1103.2541},
 primaryClass = {astro-ph.EP},
       adsurl = {https://ui.adsabs.harvard.edu/abs/2012ApJS..201...15H}
}

@ARTICLE{2006AcA....56....1G,
       author = {{Gould}, A. and {Dorsher}, S. and {Gaudi}, B.~S. and {Udalski}, A.},
        title = "{Frequency of Hot Jupiters and Very Hot Jupiters from the OGLE-III Transit Surveys toward the Galactic Bulge and Carina}",
      journal = {\actaa},
         year = 2006,
        month = mar,
       volume = {56},
        pages = {1-50},
          doi = {10.48550/arXiv.astro-ph/0601001},
archivePrefix = {arXiv},
       eprint = {astro-ph/0601001},
 primaryClass = {astro-ph},
       adsurl = {https://ui.adsabs.harvard.edu/abs/2006AcA....56....1G}
}

@ARTICLE{2013MNRAS.433..889K,
       author = {{Kov{\'a}cs}, G{\'a}bor and {Hodgkin}, Simon and {Sip{\H{o}}cz}, Brigitta and {Pinfield}, David and {Barrado}, David and {Birkby}, Jayne and {Cappetta}, Michele and {Cruz}, Patricia and {Koppenhoefer}, Johannes and {Mart{\'\i}n}, Eduardo L. and {Murgas}, Felipe and {Nefs}, Bas and {Saglia}, Roberto and {Zendejas}, Jesus},
        title = "{A sensitivity analysis of the WFCAM Transit Survey for short-period giant planets around M dwarfs}",
      journal = {\mnras},
         year = 2013,
        month = aug,
       volume = {433},
       number = {2},
        pages = {889-906},
          doi = {10.1093/mnras/stt571},
archivePrefix = {arXiv},
       eprint = {1304.1399},
 primaryClass = {astro-ph.EP},
       adsurl = {https://ui.adsabs.harvard.edu/abs/2013MNRAS.433..889K}
}

@ARTICLE{2016MNRAS.459.2275R,
       author = {{Rein}, Hanno and {Tamayo}, Daniel},
        title = "{Second-order variational equations for N-body simulations}",
      journal = {\mnras},
         year = 2016,
        month = jul,
       volume = {459},
       number = {3},
        pages = {2275-2285},
          doi = {10.1093/mnras/stw644},
archivePrefix = {arXiv},
       eprint = {1603.03424},
 primaryClass = {astro-ph.EP},
       adsurl = {https://ui.adsabs.harvard.edu/abs/2016MNRAS.459.2275R}
}
\bibliographystyle{aasjournal}

\end{document}